\documentclass[fleqn]{2023SCGE-ar}
\usepackage{tikz}
\usepackage{graphicx} 
\usepackage{subcaption}

\tikzstyle{process} = [rectangle, minimum height=1cm, text centered, draw=black, fill=white]
\tikzstyle{startstop} = [rectangle, rounded corners, minimum width=2cm, minimum height=1cm,text centered, draw=black, fill=white]
\tikzstyle{arrow} = [thick,->,>=stealth]

\usepackage{xcolor}
\usepackage{orcidlink}

\begin{document}

\ensubject{subject}

\ArticleType{Article}
\Year{2025} 
\Month{xxxx} 
\Vol{xx} 
\No{x}  
\DOI{??}
\ArtNo{xxxxxx} 
\ReceiveDate{March 8, 2025}  
\AcceptDate{xxxx xx, xxxx}

\title{Lunar Orbital VLBI Experiment: 
\\ motivation, scientific purposes and status} %
  

\author[1]{Xiaoyu Hong\orcidlink{0000-0002-1992-5260}}{{xhong@shao.ac.cn}}
\author[2]{Weiren Wu}{}
\author[1]{Qinghui Liu}{}%
\author[3]{Dengyun Yu}{}
\author[4]{Chi Wang}{}
\author[1]{Tao Shuai}{}%
\author[1]{Weiye Zhong}{}%

\author[1]{\\Renjie Zhu}{}
\author[1]{Yonghui Xie}{}%
\author[5]{Lihua Zhang}{}%
\author[5]{Liang Xiong}{}%
\author[2]{Yuhua Tang}{}%
\author[4]{Yongliao Zou}{}%
\author[6]{Haitao Li}{}%

\author[1]{\\Guangli Wang}{} %
\author[7]{Jianfeng Xie}{}%
\author[4]{Changbin Xue}{}%
\author[4]{Hao Geng}{}%
\author[1]{Juan Zhang}{}%
\author[1]{Xiaojing Wu}{}%
\author[1]{Yong Huang}{}%

\author[1]{\\Weimin Zheng\orcidlink{0000-0002-8723-8091}}{}%
\author[1]{Lei Liu}{}%
\author[1]{Fang Wu}{}%
\author[1]{Xiuzhong Zhang}{}%
\author[1]{Tao An\orcidlink{0000-0003-4341-0029}}{}%
\author[1]{Xiaolong Yang\orcidlink{ 0000-0002-4439-5580}}{}%
\author[1]{Fengxian Tong}{}%

\author[1]{\\Leonid I. Gurvits\orcidlink{0000-0002-0694-2459}}{}%
\author[8]{Yong Zheng}{}%
\author[1]{Minfeng Gu\orcidlink{0000-0002-4455-6946}}{}%
\author[9]{Xiaofei Ma}{}%
\author[1]{Liang Li}{}%
\author[1]{Peijia Li\orcidlink{ 0000-0002-7230-8093}}{}%
\author[1]{Shanshan Zhao}{}
\author[1]{\\Ping Rui}{}%
\author[5]{Luojing Chen}{}%
\author[9]{Guohui Chen}{}%
\author[7]{Ke Li}{}%

\author[1]{Chao Zhang}{}%
\author[1]{Yuanqi Liu}{}%
\author[1]{Yongchen Jiang}{}%
\author[1]{Jinqing Wang}{}%
\author[1]{\\Wenbin Wang}{}%
\author[1]{Yan Sun}{}%
\author[10]{Longfei Hao\orcidlink{0009-0002-4392-2040}}{}%
\author[11]{Lang Cui\orcidlink{0000-0003-0721-5509}}{}%
\author[1]{Dongrong Jiang}{}%
\author[1]{Zhihan Qian}{}
\author[1]{Shuhua Ye}{}%

\AuthorMark{X.Y. Hong}

\AuthorCitation{X. Y. Hong, et al}

\address[1]{Shanghai Astronomical Observatory, Chinese Academy of Sciences, Shanghai 200030, China}
\address[2]{Lunar Exploration and Space Engineering Center of the China National Space Administration, Beijing 100048, China}
\address[3]{China Aerospace Science and Technology Corporation, Beijing 100048, China}
\address[4]{National Space Science Center,  Chinese Academy of Sciences, Beijing 100190, China}
\address[5]{DFH Satellite Co.,Ltd., Beijing 100094,  China}
\address[6]{Beijing Institute of Tracking and Telecommunications Technology, Beijing 100094, China}
\address[7]{Beijing Aerospace Control Center, Beijing 100094,China}
\address[8]{Information Engineering University, Zhengzhou 450001, China}
\address[9]{Xian Institute of Space Ratio Technology,Xian,710100,China}
\address[10]{Yunnan Observatory, Chinese Academy of Sciences,Kunming 650011, China}
\address[11]{Xinjiang Astronomical Observatory, Academy of Sciences, Urumqi 650011, China}
\abstract{
The Lunar Orbital VLBI Experiment (LOVEX) is a scientific component of the Chinese Lunar Exploration Project (CLEP) Chang'E-7. The spaceborne component of LOVEX is implemented onboard the relay satellite QueQiao-2, which was launched on 2024 March 20, and later placed into an elliptical selenocentric orbit. The LOVEX-specific payload consists of an X-band cryogenic receiver, a hydrogen maser frequency standard, and VLBI data formatting and acquisition electronics. Several components of the QueQiao-2 nominal onboard instrumentation, such as the 4.2-meter antenna, the data storage device, and the downlink communication system, contribute to the overall spaceborne VLBI instrumentation. This allows us to form a space radio telescope capable of co-observing with Earth-based radio telescopes in VLBI mode. In this space VLBI system, the length of the baseline extends up to approximately 380,000 km. This paper presents the LOVEX scientific objectives, architecture, instrumentation, pre-launch tests, in-flight verification and calibration, and the first in-flight detections of interferometric response (''fringes'') achieved through observations of the quasar AO\,0235$+$164 and the Chang'E-6 orbital module, positioned at the Sun-Earth Lagrange point L2. These initial results demonstrate the successful performance of LOVEX, verifying its capability for both astronomical and spacecraft tracking observations at ultra-long VLBI baselines.
}


\maketitle

\begin{multicols}{2}
\section{Introduction}
\label{section1}

Very Long Baseline Interferometry (VLBI) is an astronomical technique distinguished by its record-high angular resolution (approximated by the ratio of the observing wavelength to the baseline projection to the picture plane,  $\phi = \lambda/B$) exceeding that of any other experimental technique. Introduced in the second half of the 1960s,  
this technique enables studies of celestial radio sources with sub-milliarcsecond (nano- and sub-nano radian) angular resolution (astrophysics applications)
as well as obtaining most accurate celestial coordinates of compact radio sources 
(astrometric applications) 
and terrestrial coordinates of radio telescopes 
(geodetic applications) 
\cite[Chapter~1, 9 to 12]{TMS-2017}. Modern Earth-based VLBI arrays operate with baselines comparable to the Earth diameter.  
The only way to further sharpen the angular resolution is to create a VLBI system with a baseline longer than the Earth diameter, thus placing at least one telescope in space, i.e., to create a Space VLBI (SVLBI) system.

Over the past half a century, there have been many attempts to create a 
SVLBI system -- see \cite{LIG-2020,LIG-2023} and references therein for a brief review of the history of SVLBI. To date, three SVLBI projects have been implemented. The first operational SVLBI was a demonstration of interferometry with baselines longer than the Earth diameter, the Orbital VLBI experiment conducted with the NASA’s geostationary Tracking and Data Relay Satellite System (TDRSS) in 1986 \cite{TDRSS-1986}. 

The first dedicated SVLBI system was implemented as the Japanese-led international VLBI Space Observatory Program (VSOP) consisting of the Highly Advanced Laboratory for Communication and Astronomy (HALCA) satellite and a global network of Earth-based radio telescopes \cite{VSOP-1998Sci}. The mission was operational in the period 1997--2003. The spaceborne component of the mission, HALCA, was equipped with a deployable antenna with an effective diameter of 8~m, a set of three radio astronomy receivers for observations at  18, 6 and 1.35~cm wavelengths,
and VLBI backend and data handling electronics. The VSOP mission conducted scientific investigations of many dozens of extragalactic radio sources (mostly –– Active Galactic Nuclei, AGN) at wavelengths 6 and 18~cm, and hydroxyl OH masers at 18~cm with baselines up to three times longer than the Earth diameter.
The VSOP mission produced a set of cutting-edge scientific results in continuum (on AGN) and OH maser studies summarised in \cite{SVLBI-2000, ASPC-402-2009} and references therein.

The second dedicated SVLBI mission was the Russian-led RadioAstron \cite{NSK+2013RA}. The 10-m spaceborne telescope of this mission was deployed onboard the Spektr-R spacecraft placed on a high-eccentricity and evolving orbit with the apogee reaching at times the value up to
about 350,000~km. It was equipped with four dual-polarisation radio astronomy receivers for observations at 92, 18, 6 and 1.3~cm.
The RadioAstron mission successfully operated in the period 2011--2019. At the time of this writing, its science data processing is still ongoing. 
But it is clear that the RadioAstron SVLBI mission has made a major impact on studies of AGN (e.g., \cite{Kovalev+2020AdSpR, Giovannini+2018NatAs, Fuentes+2023NatAs}), galactic and extragalactic masers \cite{Sobolev+2018IAUS}, pulsars and interstellar medium \cite{Smirnova+2014ApJ, Popov+2020ApJ}, as well as fundamental physics applications \cite{Litvinov+2018PhLA, Nunes+2023CQGra}. 

The Chinese VLBI Network (CVN) participated in some SVLBI observations as a part of international Earth-based VLBI networks. VLBI technique in its near-field (nfVLBI) modification has been exploited in support of orbit determination (OD) for the Chinese Lunar Exploration Project (CLEP, Chang'E-1 to Chang'E-6) and Mars Exploration (Tianwen-1) missions. The real-time VLBI system with Delta-Differential One-way Ranging ($\Delta$DOR) and same beam VLBI techniques have also been developed in synergy with nfVLBI for OD purposes and contributed to the success of the mentioned missions \cite{Hong-2020,Dong-2018,Qian-Li-2012}. SVLBI studies in China were initiated in 2012. The primary focus of this study was the development of a millimeter-wavelength VLBI array \cite{Hong-2014,An+2020AdSpR}. Significant progress was also made in the research and development of key technologies essential for SVLBI missions.

The continuing development of the China's deep space exploration program led to new requirements for the interplanetary probe's orbit determination. It was considered worthwhile to investigate the OD potential of VLBI observations of deep space probes using a baseline comparable to the Earth--Moon distance as one of the attractive options. Such an application of the VLBI technique for OD purposes is based on the extensive contributions of the CVN to the orbit determination of China's deep space exploration missions, as well as preliminary studies of Space VLBI mentioned above. After multiple demonstrations and competitive evaluation of various projects, a proposal for the Lunar Orbital VLBI Experiment (LOVEX) from Shanghai Astronomical Observatory (SHAO) was approved. The essence of the proposal was to combine the interests of SVLBI with the nominal QueQiao-2 communication hardware (the 4.2-m antenna and other equipment) and several VLBI-specific devices. In addition to the OD experiments, LOVEX will conduct astronomical observations of natural compact celestial radio sources on the ultra-long baseline between Earth and QueQiao-2 on the selenocentric orbit. 

The current paper presents the LOVEX concept, the major related technological developments, its science drivers, preparation for in-flight tests and science operations. The paper is organized as follows. Section~\ref{s:syst-des} describes the LOVEX system design and payload. Section~\ref{s:lovex-sci} presents LOVEX science objectives. The Earth-based infrastructure of LOVEX is presented in section~\ref{s:erth-infra}. Section~\ref{s:tsts} describes the pre-launch and in-flight tests of LOVEX. Section~\ref{s:svlbi-tst} presents the first interferometric fringes obtained on the baseline between Earth and selenocentric orbit. Section~\ref{s:cncl} summarizes the paper and offers brief conclusions.

\section{LOVEX system design and payload}
\label{s:syst-des}

\subsection{System design concept}
\label{ss:cncpt}

LOVEX is conceived as a valuable \textit{ad hoc} addition to the main objective of the CLEP. LOVEX utilises two sets of instrumentation: the nominal QueQiao-2 payload, designed primarily to support the relay function for  the CLEP, and the VLBI-specific devices and systems. Thus, the design concept of LOVEX aims at maximising utilisation of the equipment of the QueQiao-2 relay satellite. 
 
\begin{figure}[H]
\centering
  \includegraphics[width=\columnwidth]{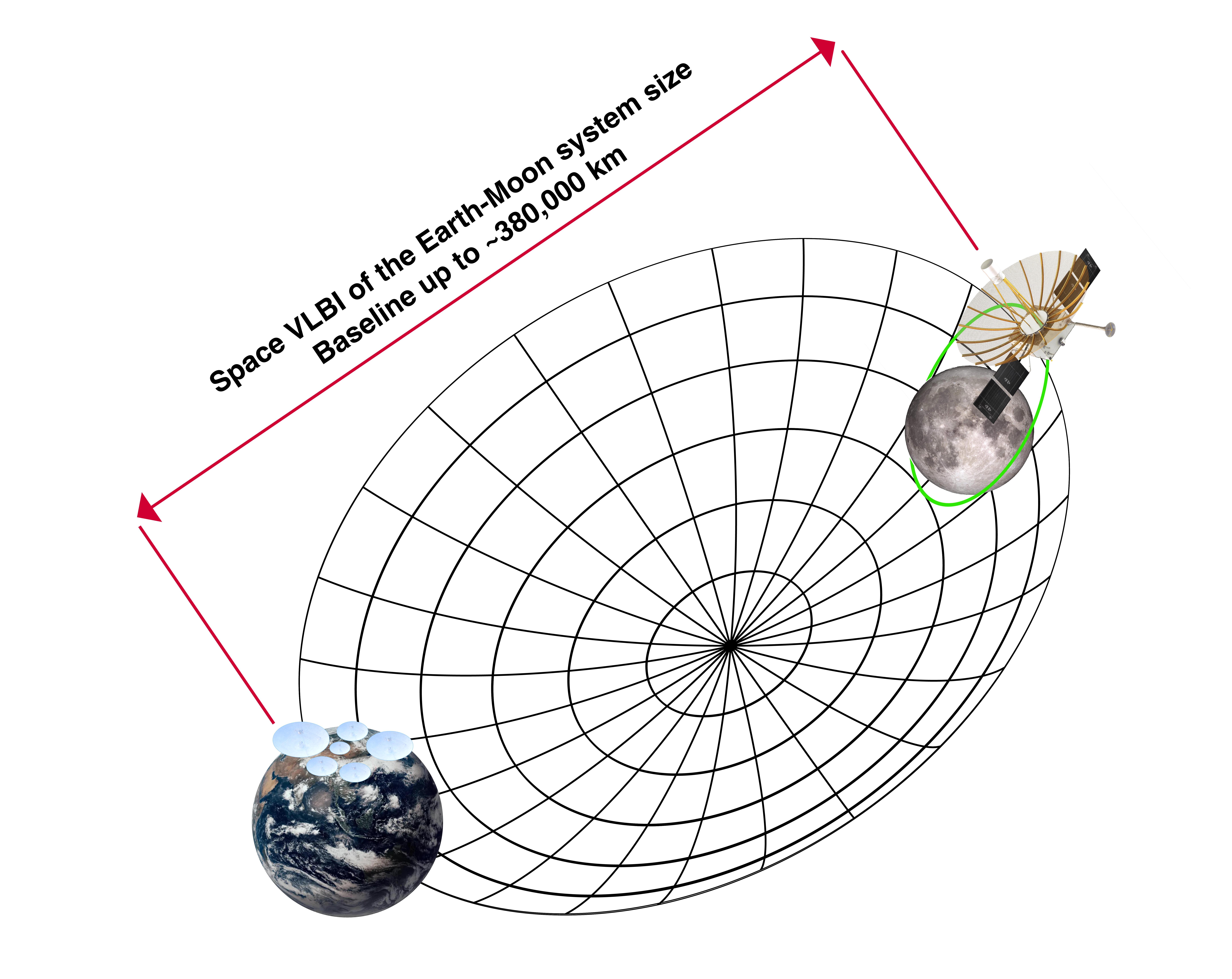}
 \captionsetup{justification=raggedright}
  \caption{A schematic configuration of LOVEX with its spaceborne radio telescope on the QueQiao-2 relay satellite on a selenocentric orbit. Together with Earth-based radio telescopes, this VLBI system has a baseline length comparable to the Earth-Moon distance, 380,000~km.
  } 
\label{fig:LOVEX-sch}
\end{figure}

LOVEX combines in  an orbital element, including the parabolic antenna and other instrumentation of 
onboard QueQiao-2 and a VLBI network of Earth-based radio telescopes. The main objective of LOVEX is to demonstrate the feasibility of VLBI detections on baselines comparable to the distance Earth--Moon in observations of natural celestial radio sources and deep space mission spacecraft (Fig.~\ref{fig:LOVEX-sch}).

\begin{figure}[H]
\centering
  \includegraphics[width=\columnwidth]{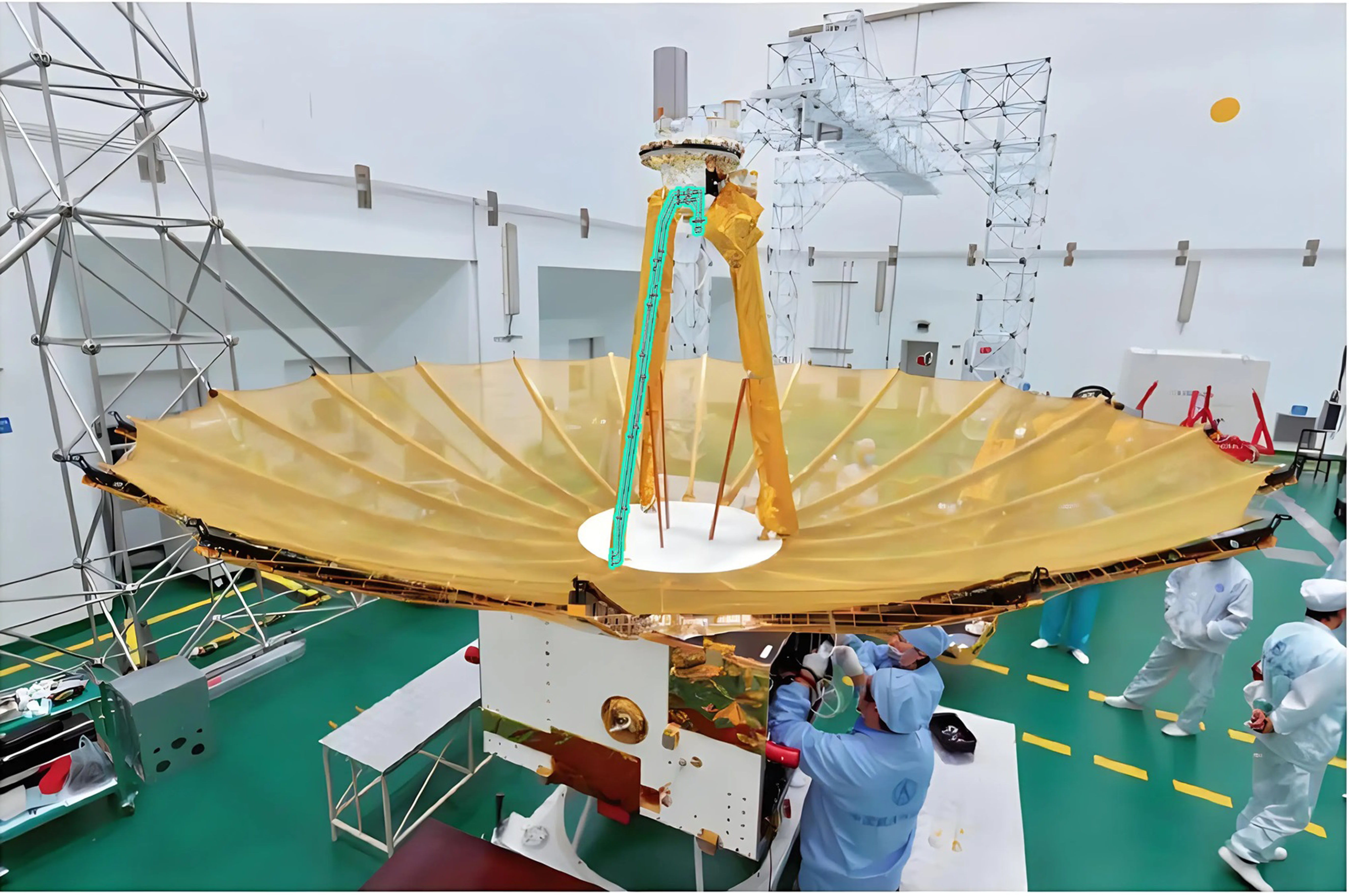}
 \captionsetup{justification=raggedright}
  \caption{The QueQiao-2 satellite at the assembly facility. The 4.2-m parabolic antenna is equipped with a dedicated LOVEX feed and wave guide (shown in bright green).}
\label{fig:queiao}
\end{figure}

VLBI-specific payload onboard QueQiao-2 covers only those functionalities, which are needed for VLBI observations.
Due to the \textit{ad hoc} character of LOVEX, its specific payload onboard QueQiao-2 must fit within the mass limit of 35~kg and power consumption less than 300~W. Fig.~\ref{fig:queiao} shows the QueQiqo-2 satellite at the assembly facility. 

\begin{figure}[H]
\centering
 \includegraphics[width=\columnwidth]{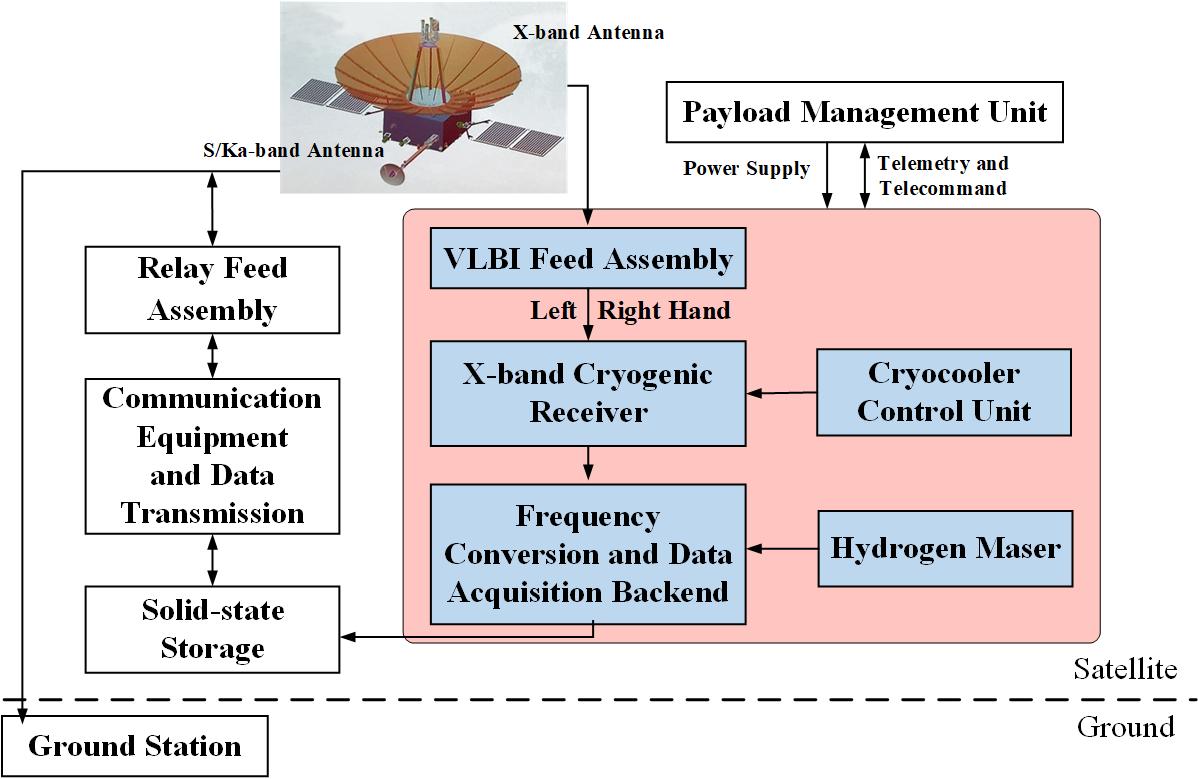} 
  \captionsetup{justification=raggedright}
  \caption{The LOVEX block-diagram. VLBI-specific payload is shown by blue boxes enclosed within the beige rectangle.}
 \label{fig:blc-dgr}
\end{figure}

Fig.~\ref{fig:blc-dgr} represents the block-diagram of the QueQiao-2 onboard instrumentation which addresses both functions of the satellite, the lunar probes' data relay and LOVEX VLBI observations. 

\subsection{QueQiao-2 instrumentation involved in LOVEX}
\label{ss:queq-2}

The QueQiao-2 satellite (Fig.~\ref{fig:queiao}) was designed primarily to operate as a radio relay component of the Phase-4 of CLEP \cite{QQ2,QueQiao-2018}. It was launched by a Long March~8 space rocket from the Wenchang spaceport on Hainan island on 20 March 2024. Four days later, it was placed on the operational selenocentric retrograde orbit with the periselene about 300~km, aposelene about 160,000~km, inclination $\sim$118.2$^\circ$, and period $\sim$24~hr. This so-called ``frozen'' orbit is sufficiently stable and is possessing the capacity to provide reliable relay communication services to multiple lunar exploration missions \cite{QQ2}.

The QueQiao-2 satellite is based on the CAST 2000 bus\footnote{\url{https://www.cast.cn/english/news/3295}, accessed 2024.11.04.}. 
It is equipped with a 4.2~m main deployable parabolic antenna and a 0.6~m Ka 
communication antenna. 

The 4.2-m parabolic antenna has a fixed mounting on the satellite (see Fig.~\ref{fig:queiao}). The antenna was folded during the launch and deployed in the beginning of the Earth-Moon cruise phase. The antenna is optimised for operations at X-band (8.4~GHz) and serves as the main component of the LOVEX radio telescope. 

A LOVEX-dedicated feed was added to the feed assembly of the 4.2-m antenna. This special feed is mounted at a 1.6$^{\circ}$ offset from the communication (relay) feed. The LOVEX feed is linked with the receiver with a waveguide and cable as shown in green in Fig.~\ref{fig:queiao}. The 0.6-meter diameter S/Ka dual-band (2.3 and 26~GHz)  parabolic antenna supports data downlink to Earth-based stations. The satellite is equipped with a data storage with the capacity of 4~Tbit \cite{QQ2}. The on-board  data storage system and the Ka-band data transmission system with 8PSK modulation and bit rate of 500~Mbps are used for both Lunar probes' data relay and raw VLBI data downlink to an Earth-based receiving station.

LOVEX relies on the QueQiao-2 service systems, such as power supply, command and control, attitude control and orbit determination, data acquisition and data transmission, time synchronization system, and scientific payload management system. 

\begin{table}[H]
\caption{. \,\,\, Mass of LOVEX-specific payload components.}
\tabcolsep 2pt 
\begin{center}
\begin{tabular}{lr}
\toprule
Item   & Mass \\
      & (kg)      \\ \hline
 X-band cryogenic receiver & 7.089  \\
 Cryocooler control unit   & 2.725  \\
 Frequency conversion $\&$ data acquisition backend  & 4.175  \\
 Hydrogen maser           & 13.710 \\
 Cable assembly           & 1.994  \\
 VLBI feed                & 2.300  \\ 
 VLBI waveguide           &  0.784  \\  \hline 
 Total                    &  32.777  \\  
\bottomrule
\end{tabular}
\end{center}
\label{tab:mass}
\end{table}

\subsection{LOVEX-specific payload}
\label{ss:pyld}

The LOVEX-specific payload onboard the QueQiao-2 satellite included several physically independent components: VLBI feed and waveguide assembly, VLBI X-band receiver (cryogenic electronics unit, cryocooler, cryocooler control unit), VLBI frequency conversion and data acquisition backend,  Hydrogen maser, and cables.

The LOVEX-specific payload was subject to strict mass and power budget limits. The total mass of the LOVEX specific payload onboard QueQiao-2, including the VLBI feed waveguide and cables, is 32.777~kg (see Table~\ref{tab:mass}), and the nominal operational power consumption is about 220~W. 

The X-band cryogenic receiver was designed to employ a pulse tube cryocooler to cool its low-noise amplifier (LNA) to a physical temperature to $-138.2^{\circ}$~C to obtain a low receiver noise temperature of 42.3~K (Table~\ref{tab:param}), enabling us to get an ultra-high sensitivity. The receiver  operates within the 8.1$-$9.0~GHz frequency range in both left-hand (LHCP) and right-hand (RHCP) circular polarisation channels. The space-qualified cryogenic receiver of LOVEX, including the cryogenic LNA and all other components, was the first such device developed and produced in China \cite{Zhong+LOVEX-2025}.

\begin{table}[H]
\caption{. \,\,\,\,Major parameters and specifications of LOVEX space-borne instrumentation.} 
\tabcolsep 2pt 
\begin{tabular}{lc}
\toprule
Item & Parameter(s)  \\\hline 

Observing band                  & X-band (8~GHz)              \\
Instantaneous bandwidth         & 8.1 -- 9.0~GHz              \\
Polarization                     & Dual circular               \\ \hline
Antenna diameter                & 4.2~m                       \\
Antenna efficiency ($\eta_a$) &  35$\%$                 \\ 
Pointing precision              &  $\sim 0.1^{\circ}$         \\
Cryo-receiver noise temperature $(T_{rec.})$   &  $\sim 42.3$~K  \\              
System temperature($T_{sys}$)   & $\sim 89.3\pm5$~K      \\ \hline
Observing bandwidth (selectable)            & 64,128,256,512MHz       \\
Maximum sampling rate           & 2048 Msps                   \\
Storage capacity                & 4 Tbit                      \\
Storage read-in data rate       & 2048 Mbps                  \\
Data transfer speed             & 500 Mbps                 \\  \hline
H-maser output frequency        & 10~MHz             \\
H-maser Allan Deviation         & $8\times 10^{-14}$ at 100~s  \\ 
                                & $3.6\times 10^{-14}$ at 500~s  \\ \hline
LOVEX-specific payload mass     &     32.78~kg              \\
\,\,\,\, (incl. feed, waveguide \& cables) &   \\
Power consumption (average)     &  $\sim 220$ W   \\ \hline
Design lifetime                 & $\geq 8$~yr      \\
Total observing time            & $\geq 1$~yr      \\   \hline
\bottomrule
\end{tabular}
\label{tab:param}
\end{table}

The data acquisition backend integrated with a frequency conversion module follows the receiver. It converts the radio frequency (RF) signals from  8.1$–$9.0~GHz range to the range of intermediate frequency (IF) 0.1$-$1.0~GHz. Then the backend performs digitization and signal processing to generate raw VLBI data stream, with the clock marks provided by a Hydrogen maser. The sampling rate reaches up to 2048 Msps, enabling the maximum observing bandwidth 512~MHz (Table~\ref{tab:param}). It connects to the storage devices via a high-speed serial bus which can operate with the speed of up to 4096~Mbps\cite{Zhong+LOVEX-2025}.

Various types of Hydrogen maser frequency standards have been developed at the Shanghai Astronomical Observatory for operations on the ground and in space. Due to the strict mass and power limitations, a passive Hydrogen maser was designed and built for the LOVEX project. A crystal oscillator is locked on the hydrogen maser signal and generates a highly-stable reference 10 MHz harmonic. It represents state-of-the-art technology, and its frequency stability meets the mission requirements, as shown in Table~\ref{tab:param} \cite{Zhong+LOVEX-2025}.  It’s the first passive Hydrogen maser in space-borne astronomy.

\begin{figure}[H]
\centering
  \includegraphics[width=\columnwidth]{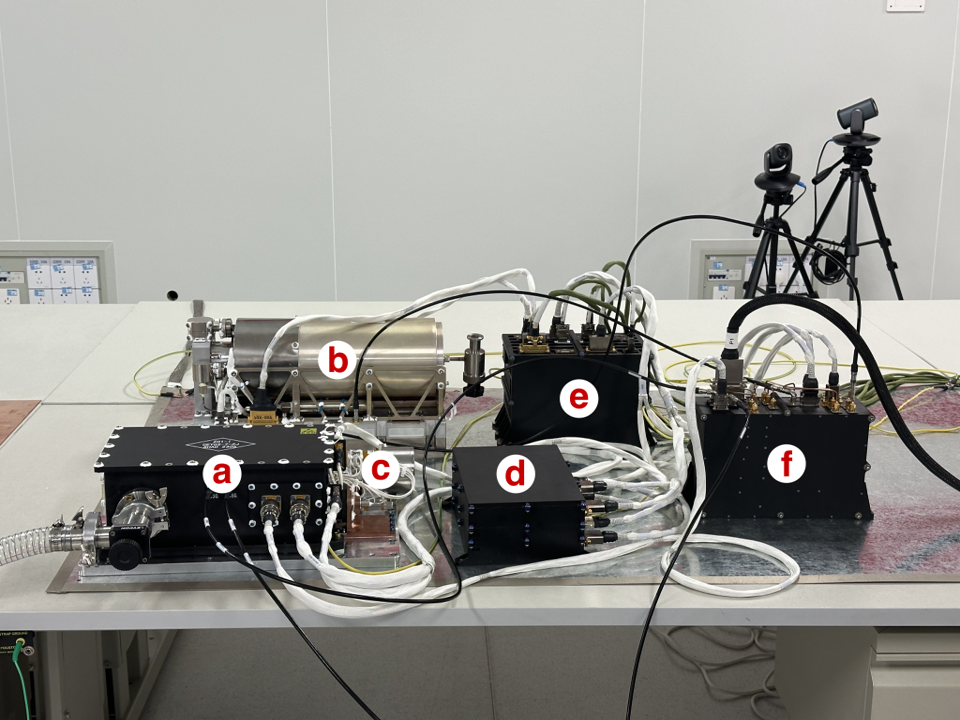} 
  \captionsetup{justification=raggedright}
  \caption{The desktop integrated test of the LOVEX flight models. (a) X-band cryogenic receiver, (b) Hydrogen maser, (c) cryocooler, (d) cryocooler control unit, (e) payload management unit, and (f) frequency conversion and data acquisition backend.}
\label{fig:l-pld}
\end{figure}

A detailed description of all payload components is given in \cite{Zhong+LOVEX-2025}. Fig.~\ref{fig:l-pld} shows the desktop integrated test of the LOVEX flight models.

Four models of the VLBI X-band cryogenic receiver, frequency conversion and data acquisition backend, and hydrogen maser have been developed and tested: prototype, electrical model (EM), qualification model (QM) and flight model (FM). Each model has undergone comprehensive electrical performance testing and participated in satellite payload system joint testing. In addition, the qualification model has successfully completed the qualification-level mechanical tests, thermal tests, and electromagnetic compatibility (EMC) tests; The flight model has undergone acceptance-level mechanical and thermal tests. 

The signal from the observing target containing both senses of polarisation, left circular and right circular, comes to the cryogenic X-band receiver through the 4.2-m antenna, the dedicated LOVEX feed and waveguide. The receiver includes cryo-electronic units, a vacuum dewar, and pulse tube cooler, etc., supporting highly sensitive RF frontend circuits. 

The data acquisition backend integrated with a frequency conversion module follows the receiver. It converts the RF signals to  IF signals. Then it performs digitization and digital signal processing to generate VLBI raw observation data, with the clock sourced by Hydrogen maser. Then, it transmits the data to storage devices.

The on-board data storage system is based on the solid-state memory electronics. Its capacity is 4~Tbit. The stored data are downlinked to the ground receiving station via the QueQiao-2 Ka-band transmission system.

A correlation processor able to process VLBI data obtained on the Earth-QueQiao-2 baseline is designed and built specifically for LOVEX. The description of its hardware and software is given in \cite{ZhangJ+LOVEX-2025}.

LOVEX can observe natural celestial radio sources in dual-polarisation with the maximum bandwidth of 512~MHz per polarisation within the frequency band 8.1$-$9.0~GHz. The actual observing bandwidth is selectable and can be 64, 128, 256, and 512~MHz. At the widest bandwidth, the highest data rate of the VLBI stream is 2048~Mbps. 

Table~\ref{tab:param} summarizes major parameters of LOVEX spaceborne instrumentation.

\section{Science objectives of LOVEX}
\label{s:lovex-sci}

As described in section~\ref{s:syst-des}, 
LOVEX is an \textit{ad hoc} experiment which heavily utilises the hardware and instrumentation designed for the primary task of the QueQiao-2 satellite -- data relay service for CLEP. Therefore, the scientific objectives of LOVEX are gauged to make them compatible with other functionalities of QueQiao-2. Nevertheless, beyond demonstration of the feasibility of VLBI observations on the Earth--Moon baseline\footnote{In the LOVEX context, the term ``Earth--Moon baseline'' is used hereafter for brevity; the actual baseline connects Earth-based radio telescopes and the QueQiao-2 satellite on a selenocentric orbit.}, LOVEX is capable of addressing science objectives in two primary scientific areas: (a) studies of ultra-compact structures in natural continuum celestial radio sources, and (b) ultra-precision estimates of the state vector of deep space mission probes using nfVLBI techniques. These objectives are consistent with the science tasks defined for LOVEX as a part of the overall scientific goals of the Chang'E-7 mission \cite{Zou+2020}. 

\begin{figure}[H]
\centering
  \includegraphics[width=\columnwidth]{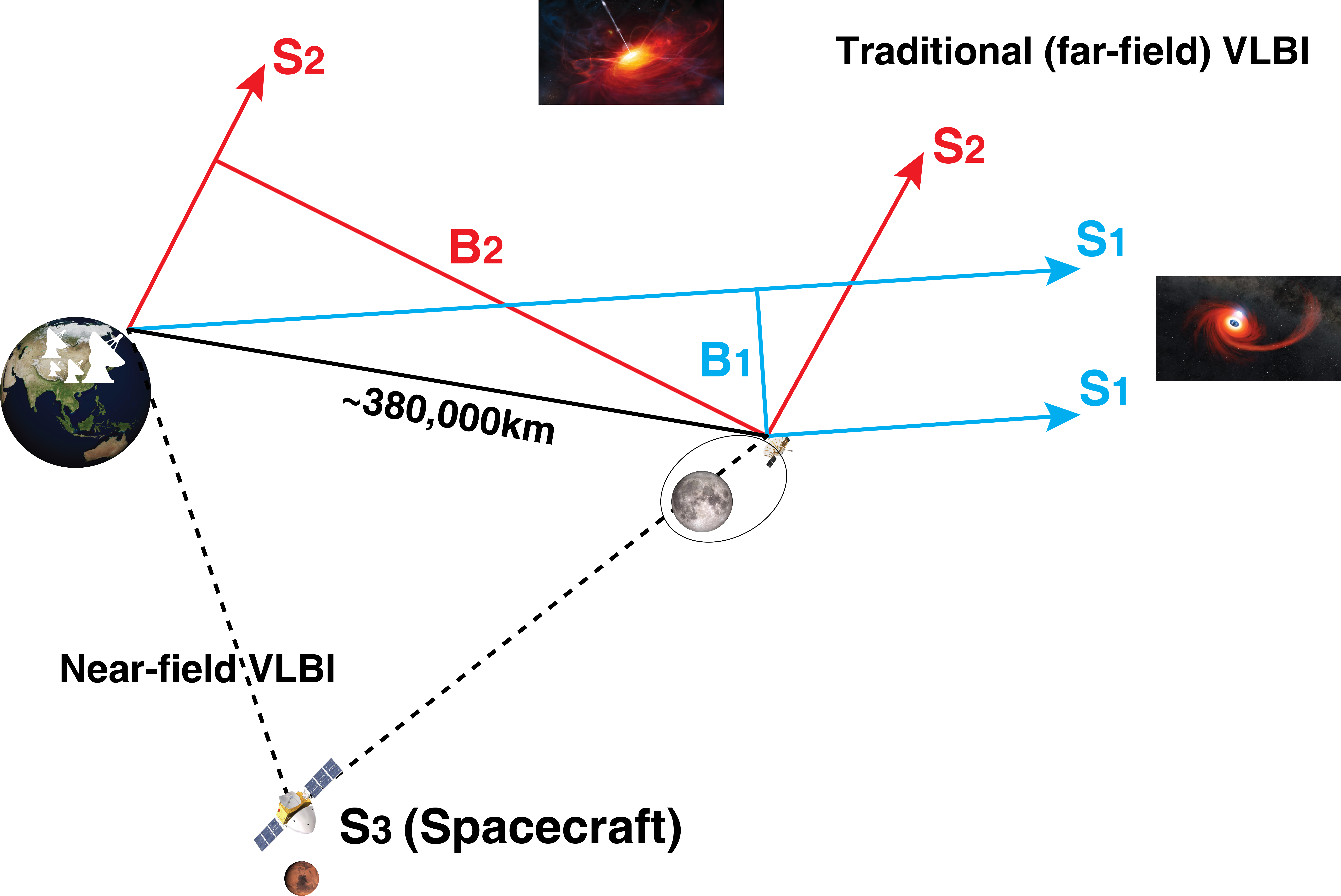}
  \captionsetup{justification=raggedright}
  \caption{Geometrical configurations of
  LOVEX observations. In observations of natural celestial radio sources S1 and S2, the interferometer, consisting of an Erath-based radio telescope and QueQiao-2, operates in ``traditional'' VLBI mode, with the target of observations located at an infinitely large distance (i.e., at a distance much larger than the Fraunhofer criterion, $B^2/\lambda$). In these two cases, the ray paths to both elements of the interferometer are parallel. The difference between the cases S1 and S2 is in the projection effect: in the S1 case, the target is located close to the direction of the baseline vector, thus providing a shorter projected baseline which are more suitable for imaging observations in combination with Earth-only baselines. In the case of S3, the target (a spacecraft) is located in the near field (closer than the Fraunhofer criterion $B^2/\lambda$), thus the ray paths to the two elements of the interferometer are not parallel. 
  }
\label{fig:orb-geom}
\end{figure}

\subsection{LOVEX interferometer geometry}
\label{ss:orb-geom}

Due to the nature of LOVEX, the length of the interferometer's baseline between Earth-based telescopes and QueQiao-2 always remains close to the distance between Earth and Moon, about 380,000~km. However, the orientation of this baseline vector in three-dimensional space continuously evolves, following both the orbital motion of the Moon in the Earth--Moon system and the orbital motion of Earth in the Solar System. Fig.~\ref{fig:orb-geom} depicts the schematic geometry of the LOVEX interferometer in two observing modes: observations of a natural celestial source and a spacecraft. In the former case, the observational target (e.g., a quasar) is located at so large distance from the observer, that it is physically acceptable to consider ray paths from the target to the telescopes (QueQiao-2 and Earth-based) parallel. This is the traditional VLBI case. In this case, the target source is located at a distance significantly larger than the Fraunhofer distance, $B^{2}/\lambda$, where $B$ is a projected baseline, and $\lambda$ - an observing wavelength. For spacecraft observations, the target is located closer than the Fraunhofer distance, i.e., located in the interferometer's near field, thus creating nfVLBI. In this case, the ray paths to the VLBI telescopes are non-parallel (see \cite{PRIDE-2023} for further nfVLBI discussion).  
\begin{figure}[H]
\centering
  \includegraphics[width=0.95\columnwidth]{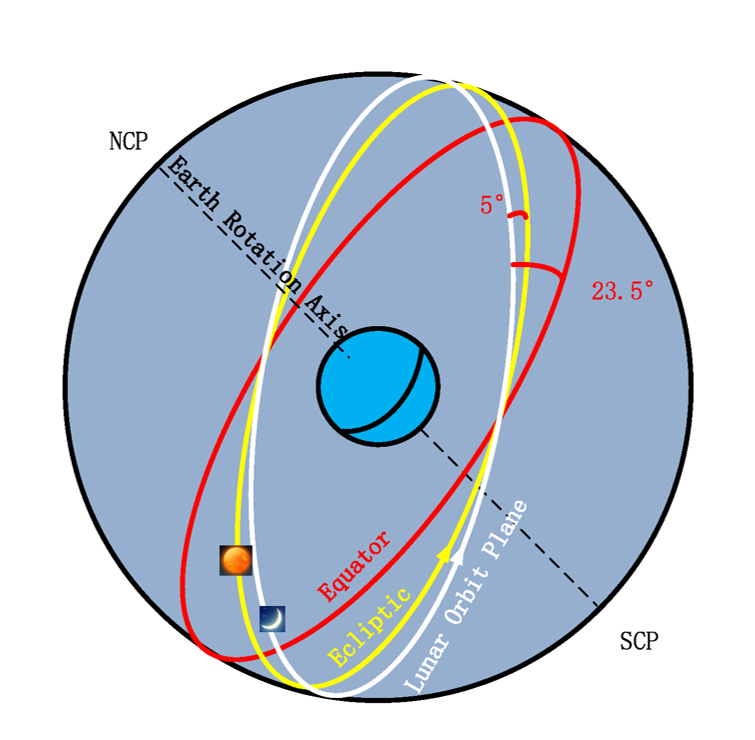}
 \captionsetup{justification=raggedright}
  \caption{Orbital geometry of the LOVEX interferometer. The QueQiao-2 on the selenocentric orbit follows the Moon trajectory (shown in white). Earth-based radio telescopes move along the ecliptic (shown in yellow). The world equator is shown in red.} 
 \label{fig:LOP}
\end{figure}
The QueQiao-2 satellite moves around the Earth along the orbit which is close to that of the Moon (Fig.~\ref{fig:LOP}). Therefore, the LOVEX interferometer baseline is always close to the Moon's orbital plane. Such geometry provides extremely long baselines, with the length comparable to the Earth--Moon distance for observing targets with celestial positions close to the perpendicular to the lunar plane orbit. Interestingly, celestial sources located close to the lunar orbital plane can be observed with a wide range of baseline lengths, at some lunar orbital phases providing projected baselines comparable to the Earth-only VLBI baselines. Benefits of this configuration are further discussed in subsection~\ref{sss:imgng}. 

\subsection{LOVEX capability analysis}
\label{ss:lovex-cpbl}

The major advantage of VLBI, making it highly attractive for astronomical studies, is the angular resolution defined by the ratio $\phi = \lambda/B$ mentioned above. For the longest terrestrial baselines comparable to the Earth's diameter at observing wavelength $\lambda \simeq 3.6$~cm (X-band communication frequency 8.4\,GHz), the sharpest achievable angular resolution is $\phi \simeq 0.7$ milliarcsecond (mas). However, numerous astrophysical studies require angular resolution at a given wavelength at least an order of magnitude sharper. This necessitates a proportional increase in baseline length by an order of magnitude.  Such extended baselines can only be achievable by placing at least one radio telescope of a VLBI system in space. LOVEX, with its longest possible baseline about 380,000~km, 
 enables observations of celestial radio sources with the angular resolution reaching $\phi \simeq 20$\,$\mu$as (0.1~nanoradian) though the practical resolution of LOVEX
will depend on the baseline projected on the image plane of a celestial radio source.

Beyond angular resolution, another important characteristic of a VLBI system is its baseline sensitivity -- an ability to detect a response to a signal from cosmic sources against the background noise. The fringe noise $S_{ij}$ for an interferometer consisting of two telescopes $i$ and $j$ is defined by the following equation (see \cite{TMS-2017}):

\begin{equation}
\label{eq:bas-sen}
    S_{ij} = \frac{1}{\eta_{s}}\sqrt{\frac{SEFD_{i}\times SEFD_{j}}{2\tau\Delta \nu}} \,\,\, ,
\end{equation}
\noindent where $\tau$ is the integration time, $\Delta \nu$ the detection bandwidth, $\eta_{s}$ a coefficient of the order of unity, and $SEFD$ (System Equivalent Flux Density) quantifies the sensitivity of an individual telescope, defined as
\begin{equation}
\label{eq:sefd}
    SEFD = \frac{2kT_{\rm sys}}{\eta_{a}A} \,\,\, .
\end{equation}
\noindent In the latter formula, $k$ denotes the Boltzmann constant, $T_{\rm sys}$ is the telescope system temperature, $A$ is the geometric area of the antenna aperture, and $\eta_{a}$ is the antenna efficiency. 
For LOVEX, the system parameters listed in Table~\ref{tab:param} lead to a nominal sensitivity value $SEFD_{\rm LX} \simeq 50,000$~Jy.

As it is clear from the equations (\ref{eq:bas-sen}) and (\ref{eq:sefd}), a higher baseline sensitivity can be achieved by larger telescope apertures ($A$), lower system temperature $T_{sys}$, longer integration time $\tau$, broader observing bandwidth $\Delta \nu$, or a combination of these factors. 

A distinctive feature of LOVEX compared to all previous SVLBI missions is its exceptionally broad bandwidth capability of 512 MHz.  
Table~\ref{tab:sensi} lists the estimated sensitivity for the LOVEX--Tianma baseline for four different VLBI data rates, assuming an integration time $\tau=300$~s. The signal-to-noise ratio (SNR) values, as an example, are calculated for a hypothetical compact celestial radio source with the flux density 50~mJy. 

\begin{table}[H]
\caption{. \,\,\, Estimated baseline 1-sigma noise and SNR of fringe detection of a 50~mJy radio source for the baseline  QueQiao-2 (LX)--Tianma (TM) with 2-bit quantization.}
\tabcolsep 2pt 
\begin{center}
\begin{tabular}{ccc}
\toprule
Bandwidth   & 1-sigma noise & SNR \\
 (MHz)      & (mJy)      &      \\
            \hline
 512 & 5.0 & 10.0 \\
 256 & 7.1 & 7.1  \\
 128 & 10.1 & 5.0  \\
 64  & 14.1 & 3.5 \\
\hline 
\bottomrule
\end{tabular}
\end{center}
\begin{tablenotes}
 \item[] VLBI system parameters: $SEFD_{\rm LX}=50,000$~Jy, $SEFD_{\rm TM}=50$~Jy, $\tau=300$~s, $1/\eta_{s}=1.75$\footnotemark[6]
\end{tablenotes}
\label{tab:sensi}
\end{table}
\footnotetext[6]{\url{https:// ivscc.gsfc.nasa.gov/IVS_AC/sked_cat/SkedManual_v2018October12.pdf}, accessed on 2025-04-15.}

\subsection{Astronomical applications of LOVEX}
\label{ss:astr-my}

VLBI offers the sharpest view of astronomical objects. It allows astronomers to study astrophysical processes with unparalleled angular resolution, thus getting into the inner regions of the objects, inaccessible to other astronomical techniques. In the context of LOVEX, astronomical applications are centered around studies of physical processes in AGN -- the most powerful ``engines'' in the Universe. The following sections outline several astronomical tasks which are consistent with the observational abilities of LOVEX. 
These investigations will not only advance our understanding of extreme astrophysical environments but also serve as performance benchmarks for future Space VLBI missions.

\subsubsection{Brightness temperature and compactness of AGN}
\label{sss:non-img}

The major novel characteristic of LOVEX is its record-long baseline extending to $\sim$380,000~km. Comparing to the longest achievable baseline between Earth-based telescopes and QueQiao-2 satellite, the baselines between Earth-based telescopes themselves are short. Thus, for celestial sources located near normal to the lunar orbit plane, any triangle consisting of projected on the image plane baselines between two Earth-based telescopes and QueQiao-2 is degenerate. This makes imaging observations on this longest baseline configuration very challenging. However, this configuration is ideally suited for non-imaging investigations of the highest brightness temperatures, $T_{B}$, of AGN cores. For a slightly resolved source, the value of brightness temperature can be represented by the following formula:
\begin{equation}
    T_{B}=\frac{2\ln{2}}{\pi k}S(1+z)\frac{\lambda^{2}}{\Theta} \propto S(1+z)B^{2}\,\, ,
    \label{eq:br-temp}
\end{equation}
\noindent where $S$ is the source's flux density, $z$ its redshift, $\lambda$ the observing wavelength, $\Theta$ the source's solid angle, and $B$ the projected baseline length. 
This formula highlights that investigations of extreme brightness temperatures require the longest achievable physical baselines rather than angular resolution.

In standard models of AGN, their radio emission predominantly originates from synchrotron processes. As shown in \cite{Kel-Pau-1969}, the inverse Compton scattering imposes a brightness temperature limit in stationary emission regime at $T_{B} \lesssim 10^{11.5}$~K. While this limit generally holds across most AGN populations, notable exceptions do exist.
To date, the highest directly measured brightness temperature (not merely a lower limit estimate of $T_{\rm B, max}$)
in AGN was reported from VSOP observations of the blazar AO~0235$+$164, reaching $T_{B}\approx 6\times 10^{13}$~K, \cite{Frey+2000}. Similarly, the RadioAstron observations of 3C\,273 indicated a brightness temperature exceeding $10^{13}$~K  \cite{Kovalev+2016-3C273}, though this extreme value may be attributed to refractive substructure effects discovered by RadioAstron \cite{Johnson+2016-3C273}.
Alternative explanations for brightness temperature exceeding the Inverse Compton limit are discussed in the literature, including anomalously high Doppler factors \cite{Rees-1967-TB}, proton synchrotron emission \cite{Kardashev-2000}, coherent emission \cite{Begelman+2005}, or transitory (non-stationary) processes \cite{Tsang+Kirk-2007}.

\begin{figure*}
\centering
  \includegraphics[width=0.95\textwidth]{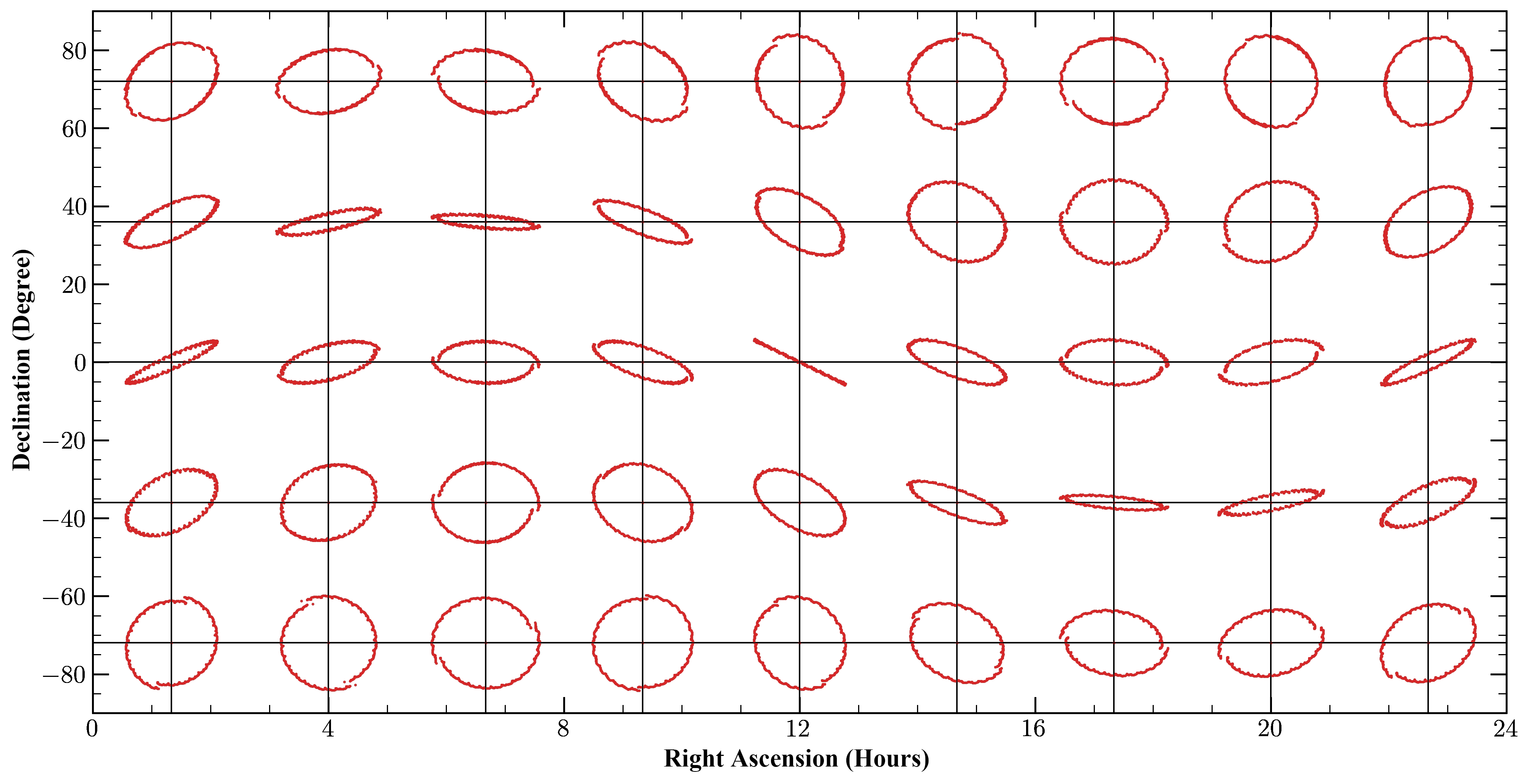}  
  \captionsetup{justification=raggedright}
  \caption{A full-sky map with examples of ideal $uv$-coverage by LOVEX over one month for sources located at crosspoints of the map's grid lines. Projected Earth-only baselines are much smaller than those between Earth-based telescopes and QueQiao-2 on a selenocentric orbit. Only for sources which are located on the sky close to the sky track of Moon, at certain times, projected baselines to QueQiao-2 are comparable to the Earth-only projected baselines (i.e., the $uv$-track passes close to the $uv$-plane origin).}
\label{fig:fsky-uv}
\end{figure*}

\begin{figure*}
\centering
  \includegraphics[width=0.95\textwidth]{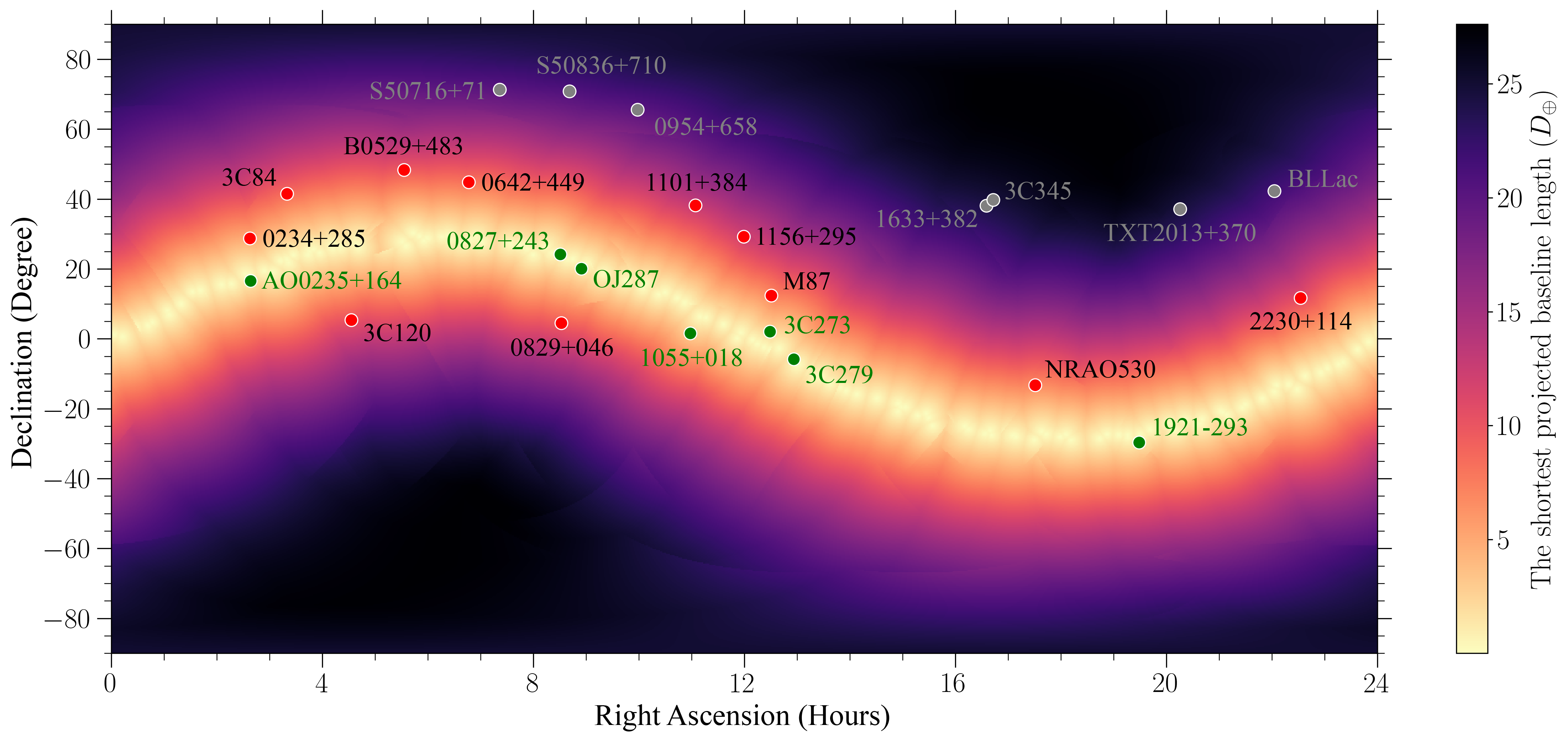}  
  \captionsetup{justification=raggedright}
  \caption{The same sky map as in Fig.~\ref{fig:fsky-uv} but with colour coding which represents the shortest achievable projected baseline length to QueQiao-2 for various source positions on the sky. The bright yellow stripe corresponds to the lunar orbital plane. This stripe indicates those celestial positions for which projected baselines to QueQiao-2 comparable to Earth-only projected baselines are possible. Sources indicated on the map and reasonably close to the median line of the bright yellow stripe might be suitable for LOVEX imaging observations. Minimum projected baseline $D_{uv}$ achievable for potential LOVEX targets: $D_{uv} < 3D_{\oplus}$ shown by green dots, $3D_{\oplus} \leq D_{uv} \leq 15D_{\oplus}$ shown by red dots, and $D_{uv} > 20D_{\oplus}$ shown by grey dots.}

\label{fig:fsky-lunar}
\end{figure*}

As is clear from the equation~\ref{eq:br-temp}, an interferometer's ability to detect high brightness temperature is proportional to the square of the projected baseline. Therefore, LOVEX should detect $T_{\rm B, max}$ exceeding maximum values measurable with Earth-based, VSOP and RadioAstron baselines by factors of approximately 1600, 200 and 2.5, respectively. Investigation of the brightness temperature limits in AGN with LOVEX would be reasonable to start from the known bright objects, of which AO~0235$+$164 is probably the best candidate. Other sources 
include 
OJ\,287, 3C\,273 and several known sufficiently strong extragalactic sources demonstrating the effect of intra-day variability (IDV), \cite{Tingay+2003var, Liu+2018var}. The IDV effect necessitates the presence of the most compact structural components in the source. 

By detecting the ultra-high brightness temperatures in AGN, LOVEX will provide crucial insights into radiation mechanisms under extreme physical conditions, potentially making transformative contributions to high-energy astrophysics. 

As it is clear from the equation \ref{eq:br-temp}, brightness temperature is de facto a measure of compactness of a source. Thus, LOVEX observations have a potential to estimate sub-milliarcsecond angular scales of compact features in radio source structures. Due to its uniquely long baseline, LOVEX can significantly advance 
non-imaging surveys of extragalactic radio sources, continuing the line of research initiated by surveys conducted by the two previous Space VLBI missions, VSOP \cite{Hirabayashi+2000PASJ} and RadioAstron \cite{Kovalev+2020AdSpR}. 

\subsubsection{Imaging LOVEX observations}
\label{sss:imgng}

In interferometric observations, signals recorded by individual telescopes are transferred into discrete samples of a two- or three-dimensional Fourier transform of the source's brightness distribution. An inverse  Fourier transform of interferometric data yields the source's brightness distribution (the source's image). In VLBI post-correlation data processing, the independent variables of this Fourier transform, amplitudes and phases, are expressed as functions of spatial frequencies, called uv-coordinates in the interferometric context. The fidelity of reconstructed images of celestial radio sources improves significantly with more complete uv-plane coverage. 

LOVEX, while not optimised for comprehensive $uv$-coverage due to its \textit{ad hoc} nature, can nonetheless produce valuable imaging data in specific geometric configurations.  Fig.~\ref{fig:fsky-uv} shows ideal $uv$-coverage patterns achievable by LOVEX for sources at different locations on the sky over a month-long observation period. 
During one month (an approximate geocentric period of the Moon), for most celestial locations, the QueQiao-2 baselines to Earth-based telescopes do not come close to the origin of the $uv$-plane. Thus, in all these configurations with very large difference in moduli of baseline projections between Earth-only and Earth--QueQiao-2 telescopes, imaging observations are impractical. However, at specific epochs for certain sources, projection effects can create more favourable $uv$-coverage for imaging purposes. 

\begin{figure*}
\centering
  \includegraphics[width=0.95\textwidth]{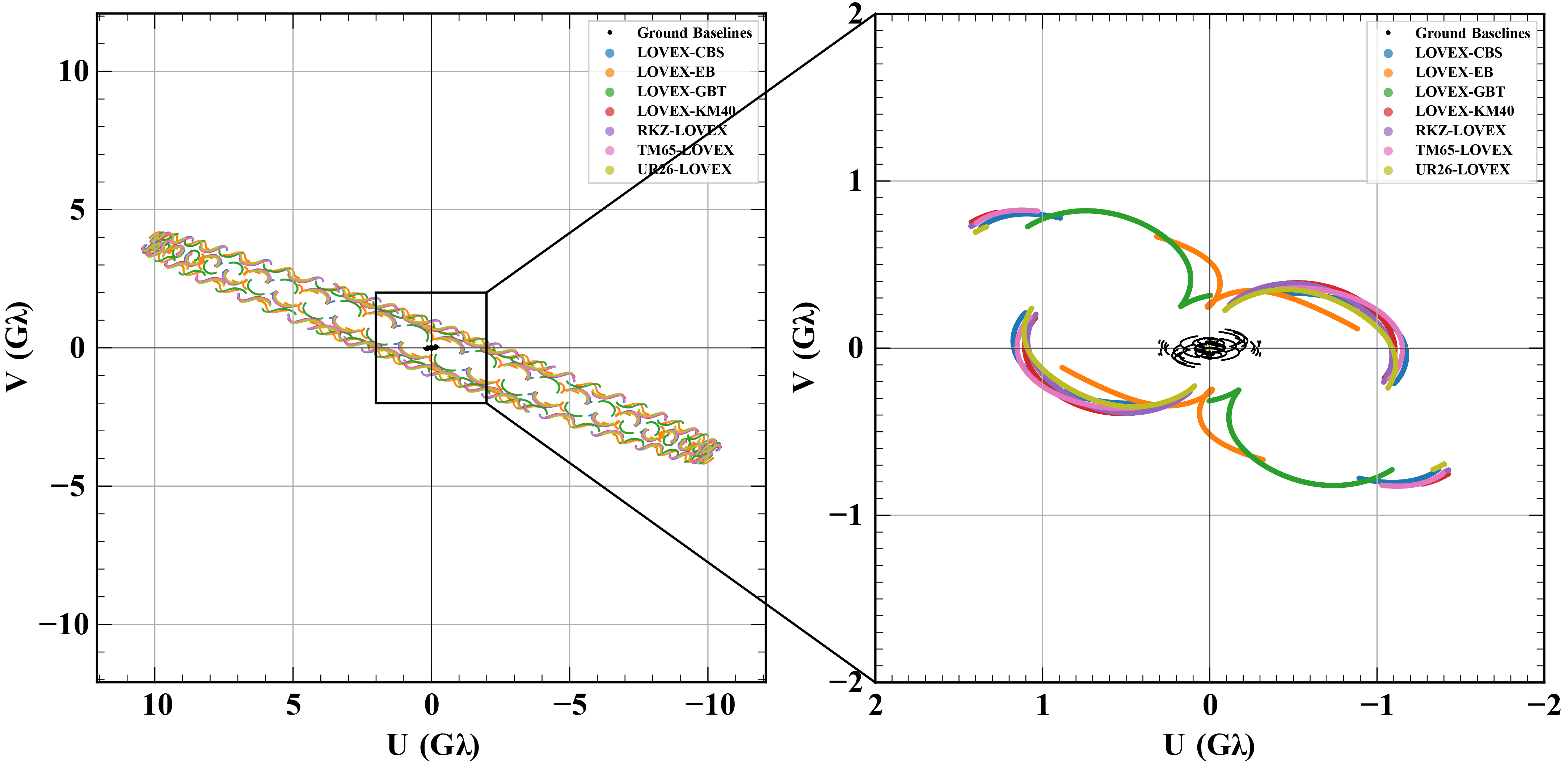}
   \captionsetup{justification=raggedright}
  \caption{An example of LOVEX $uv$-coverage in the geometrical configuration that provides projected baselines comparable to the Earth diameter for the source AO\,0235$+$164 at certain orbital phases. In this configuration, the Earth-based network consists of five Chinese radio telescopes (Tianma, Nanshan, Kunming, Changbaishan, and Shigatse, Effelsberg radio telescope (Germany) and Green Bank radio telescope (USA). The left panel shows an ideal $uv$-coverage over a 30-day-long period. The right panel shows a 12-hour-long $uv$-coverage.}
\label{fig:AO0235}
\end{figure*}

However, at some epochs for some sources, due to the projection effect, the LOVEX $uv$-coverage might be suitable for imaging. This projection effect is also illustrated in Fig.~\ref{fig:fsky-lunar}, in which several extragalactic radio sources are located in the bright yellow stripe on the sky indicating the lunar orbit. Fig.~\ref{fig:AO0235} shows such an example for the source AO\,0235+164 with a strong projection effect. In this example, projected baselines to QueQiao-2 shorter than 5~G$\lambda$ offer promising conditions for VLBI imaging synthesis. As demonstrated by RadioAstron, successful imaging with similarly structured $uv$-coverages have been demonstrated for OJ\,287 with projected baselines of up to about $6D_{\oplus}$  \cite{Cho+2024}, and for 3C\,279 with baseline extending to about $8D_{\oplus}$ \cite{Fuentes+2023NatAs}. On Fig.~\ref{fig:fsky-uv}, for some celestial sources located close to the projection of the Moon orbit, $uv$-tracks are passing close enough to the $uv$-plane origin making imaging observations feasible. 

\subsection{Towards ultra-precise astrometric applications of Space VLBI}
\label{ss:astrmtry}

VLBI technique provides the highest precision measurements of celestial coordinates of observed sources (astrometry) and terrestrial coordinates of VLBI telescopes (geodesy). In the  Earh-based astrometric VLBI applications, the positional precision for celestial radio sources is reaching tens of microarcseconds and create a foundation of the International Celestial Radio Frame (ICRF), while geodetic VLBI monitoring programs provide values of global baselines, comparable in length to the Earth's diameter, with the precision of single-digit millimeters ({\cite{Krasna+2025} and references therein). In both applications, fundamentally, the precision of measurements is defined by the diffraction phenomenon and is proportional to the expression $\lambda/B$, where $\lambda$ is the observing wavelength, and $B$ is the projected baseline. 
Both astrometric and geodetic disciplines face high-priority scientific challenges requiring precision beyond current Earth-based VLBI capabilities. A natural extension, particularly for astrometry, involves observations with baselines exceeding Earth's diameter—precisely what Space VLBI offers.

Astrometric observations with spaceborne radio telescopes will allow development of ultra-precise delay models, supported by ultra-precise determination of the interferometric baseline state vector during the entire VLBI observation, with the precision reaching fractions of the wavelength. 
While conducting such observations is a serious challenge for SVLBI instrumentation and data processing technology, their certain elements require ``live'' tests and tuning-up in real Space VLBI observations. In this sense, LOVEX, with its record-long baseline (parameter $B$ in the above expression) offers a unique platform for developing future astrometric Space VLBI applications. It is feasible to aim in such the prospective observations to the level which will improve the best achieved Earth-based VLBI astrometric figures at least by a factor of ten, thus reaching a single-digit microarcsecond precision level. Among other applications, this would facilitate potential future Space VLBI move observations aiming at measuring variations of the celestial position of compact celestial radio sources as a function of the Solar impact parameter of ray paths toward widely-separated VLBI telescopes. Such the experiments would enable verification experiments in fundamental theory of gravitation \cite{Will, Fomalont, Lambert, Bertotti}.

\subsection{Orbit determination and deep space navigation applications of LOVEX}
\label{ss:nfVLBI-LO}

LOVEX offers a valuable opportunity for advancing orbit determination methods that will enhance the dynamical modelling for future selenocentric spacecraft and provide critical navigation support for deep space missions by employing the nfVLBI techniques. This nfVLBI technique has been widely used in the Chinese deep space exploration projects for accurate state vector and orbit determination purposes \cite{Hong-2020,Qian-Li-2012}; see also \cite{PRIDE-2023} for a review of nfVLBI applications in precise spacecraft tracking. High-precision nfVLBI systems have been used extensively for orbit determination for the series of the Chinese lunar missions Chang'E-1,2,3,4,5,6 and the Martian mission Tianwen-1 \cite{Hong-2020, Liu+2022}. 

So far, all such experiments and applications have exclusively exploited Earth-based radio telescopes. LOVEX, which operates at the deep space communication frequency 8.4~GHz, is uniquely able to demonstrate for the first time nfVLBI using a spaceborne radio telescope in conjunction with an Earth-based array. These observations could be conducted in concurrence with $\Delta$DOR measurements, also demonstrating the first implementation of this technique using baselines longer than the Earth's diameter. The observations will be supported by precise state-vector estimates of the QueQiao-2 satellite. If successful, these demonstration experiments could establish foundations for future ultra-precise nfVLBI and $\Delta$DOR measurements for prospective deep space missions. The targets which could be considered for LOVEX nfVLBI observations include the Chang'E-6 spacecraft when it reaches the Earth-Sun L2 Lagrangian point, the Tianwen-1 Mars orbiter, and the Tianwen-2 asteroid sample return mission \cite{Zhang+2024TW}, which was launched on 2025 May 29. 

\section{Earth-based LOVEX infrastructure}
\label{s:erth-infra} 

In addition to an array of Earth-based radio telescopes and some QueQiao-2 service systems (briefly mentioned in subsection~\ref{ss:queq-2}), LOVEX SVLBI requires non-standard data acquisition protocol for VLBI data streams, VLBI  correlation and post-processing facilities.

\subsection{LOVEX VLBI data handling and processing}

Efficient data handling and processing are the core components of any successful VLBI experiment. These systems must acquire, buffer, and record digital VLBI data streams, transfer them to correlation centers, and perform pre-processing, correlation, and post-processing operations. The central element in this workflow is the correlator, a specialised computing system that transforms raw observational data into scientifically useful measurements.
A dedicated VLBI data correlation system has been developed specifically for LOVEX with the aim of providing correlation on the baselines between the selenocentric QueQiao-2 orbit and Earth. The system is capable of generating VLBI delay models for both the selenocentric orbit telescope and Earth-based telescopes. The data processing system also performs an initial clock offset search for the LOVEX baseline using wide-band white noise signals from extragalactic radio sources (far-field VLBI) or spacecraft (SC) transmission signals (near-field VLBI). The system can operate within  an exceptionally wide delay window [$-$10,~$+$10]~ms, a distinctive feature of this system. To accommodate various scientific objectives (astronomy and state vector determination of SC), the correlator generates output data in multiple formats optimised for different analysis workflows. Additionally, the system provides specialised outputs containing VLBI residual delay and delay rate measurements in SC nfVLBI observations. Anticipating future expansion of more Earth-based radio telescopes in future LOVEX observations, the system is designed to support the data correlation of up to 10 VLBI stations simultaneously, with a maximum bandwidth of 512~MHz and dual-polarisation per station. The correlation engine is implemented on a high-performance CPU+GPU computing cluster. Further detailed description of the LOVEX correlator is given in \cite{ZhangJ+LOVEX-2025}.

\subsection{Earth-based VLBI stations}
\label{ss:grts}

During the in-flight LOVEX VLBI tests between October 2024 and April 2025, three CVN stations participated in the observations: Tianma, Kunming, and Nanshan. Two new Chinese VLBI telescopes, Changbaishan (40~m) and Shigatse (40~m), joined LOVEX test observational campaigns in June 2025. Table~\ref{tab:coord-st} summarizes the key parameters of these Earth-based telescopes and their performance characteristics when paired with LOVEX.

\begin{table}[H]
\footnotesize
\caption{\;\;\; The Earth-based radio telescopes involved in described in this paper LOVEX operations, and their major parameters. }
\label{tab:coord-st}
\tabcolsep 4pt 
\begin{tabular}{lccccc}
\toprule
Station name  &  Diameter & Efficiency & $T_{sys}$ &SEFD & 1$\sigma$ noise$^{\ast}$ \\ 
             & (m) & ($\%$) & (K) & (Jy) & (mJy)\\ \hline

 Tianma (TM)  &   65 & 50 & 30 & 50 & 5.0 \\
 Kunming (KM) &   40 & 40 & 50 & 275& 11.7 \\
 Nanshan (UR)  &   26 & 53& 35 & 345 &13.1 \\
 Changbaishan (CB) & 40 & 66 & 26 &86 & 6.6\\
 Shigatse (ZF) &   40 & 70 & 23 &72 & 6.0\\ \hline
\bottomrule
\end{tabular}
\begin{tablenotes}
 \item[] $\ast: 1~\sigma$ noise is the baseline sensitivity to LOVEX with 512~MHz bandwidth. The related parameters are same to Table 3. 
\end{tablenotes}
\end{table}

\section{LOVEX system tests}
\label{s:tsts}

LOVEX, as an integrated functional system, has undergone several tests on the ground and in orbit. The following subsections provide a systematic overview of these validation tests. 

\subsection{Earth-based pre-launch LOVEX verification tests}
\label{s:zero-B-tst}

The end-to-end tests of the LOVEX payload started in laboratory environment from the qualification model. Among other things, these tests enabled calibration of the radio frequency bandpass, the stability parameters (Allan deviation) of the Hydrogen maser frequency standard and estimates of the system noise temperature.

In order to test the LOVEX system in ``live'' observations, the qualification model, including the cryogenic electronic unit,  the frequency conversion and data acquisition terminal, and the hydrogen maser, has been installed on the 4.5-m antenna at the Tianma telescope campus of the Shanghai Astronomical Observatory in Sheshan. Together with the Tianma 
and Nanshan radio telescopes, this test configuration alternately observed at X-band a natural celestial radio source 1741$-$038 and the $\Delta$DOR signal from the Tiannwen-1 orbiter. VLBI fringes were successfully detected on all baselines for both targets on 3 January 2024.

Fig.~\ref{fig:fri-gtest} shows the fringe detection from this test observation of the radio source 1741$-$038 on the baseline between Tianma and the 4.5-m antenna. The fringes of the main carrier of Tianwen-1 were detected on all baselines with an 8~kHz frequency resolution and a 30~s integration. A comprehensive analysis of these test observations is presented in \cite{ZhangJ+LOVEX-2025}.}
 
The flight models of all components of the LOVEX-specific payload have undergone rigorous pre-launch testing in accordance with the requirements for the Chinese aerospace products.
\begin{figure}[H]
\centering
  \includegraphics[width=\columnwidth]{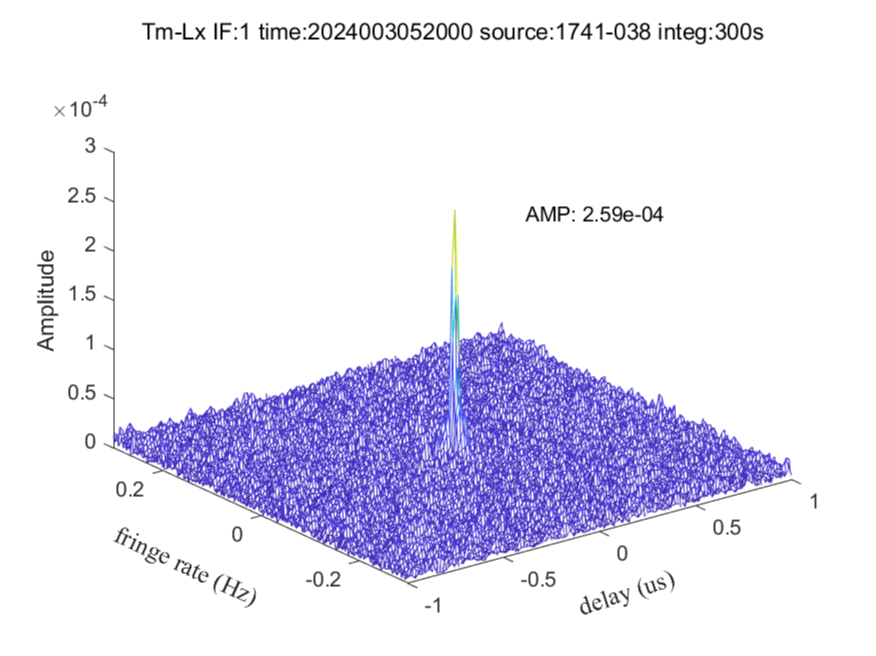}
  \captionsetup{justification=raggedright}
  \caption{Detection of the VLBI response (fringe) in the pre-launch test observation of the radio source 1741$-$038 on the baseline Tianma -- 4.5-m antenna. }
\label{fig:fri-gtest}
\end{figure}

\subsection{LOVEX in-flight verification and calibration campaign}
\label{ss:inflght-tst}

During the in-flight testing phase of LOVEX, three primary tasks have been addressed: individual instrument functional verification, system noise temperature calibration, and pointing accuracy calibration. 
On 28 June 2024, we started the individual instrument functional verification. From July to September of 2024, LOVEX performed single-dish test observations of ``cold sky'', Moon, Earth and the radio source Taurus~A  (the Crab Nebula, M1). The goal of the observations was to estimate the LOVEX instrumentation system temperature and the pointing calibration parameters. 

\subsubsection{LOVEX instrument functional verification}
\label{sss:funct-tst}

\begin{figure}[H]
\centering
  \includegraphics[width=\columnwidth]{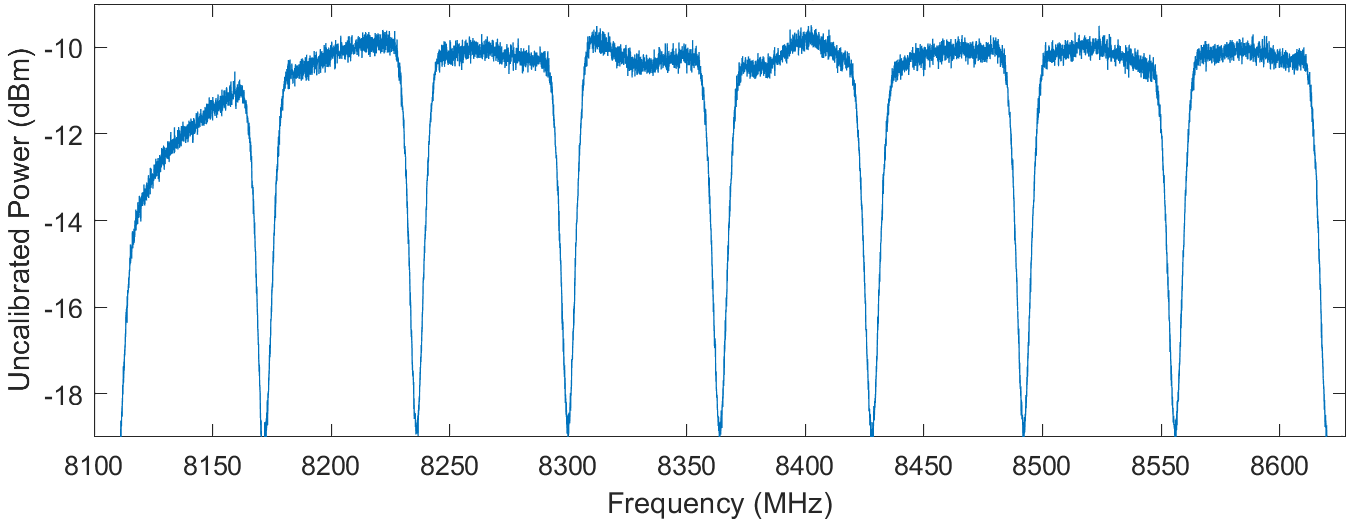}
  \captionsetup{justification=raggedright}
  \caption{Spectrum of the LOVEX passband in the 512~MHz bandwidth mode obtained during the in-flight test of 28 June 2024. The spectrum spans the frequency range 8108--8620~MHz. }
\label{fig:LOVEX-bpass}
\end{figure}

On 28 June 2024, LOVEX executed its first in-flight power-on test. Under the remote control from ground-based deep space communication stations, the payload components including the Hydrogen maser atomic clock, the cryocooler, the low-noise amplifier, and the frequency conversion/data acquisition backend were sequentially powered on. The status of all components of the payload  has been checked using data provided via the QueQiao-2 housekeeping telemetry. The test operations continued for several hours, during which the Hydrogen maser achieved a phase lock at 10~MHz reference frequency, the cryocooler demonstrated stable refrigeration performance, the low-noise amplifier maintained specified operating current and gain parameters, and the frequency conversion/data acquisition backend successfully generated standard VLBI observational data across multiple bandwidth modes. Data transmission and recording integrity have also been validated. Fig.~\ref{fig:LOVEX-bpass} demonstrates the 512-MHz-wide spectrum divided into 8 contiguous sub-channels, each 64~MHz wide. The passband (at 3~dB level) exceeds 48~MHz per sub-channel, with gaps between passbands  below 16~MHz. The lower amplitudes in the lowest frequencies of the first sub-channel are due to the sideband effects of the filter.

\subsubsection{System noise temperature calibration}
\label{sss:sys-nse}

System noise temperature, $T_{\rm sys}$, is a key component in defining the sensitivity of a radio telescope. In LOVEX, $T_{\rm sys}$ can be measured by observing a natural celestial body (e.g., Moon) , or by using an internal noise source which is installed in the signal path of the radio telescope. The nominal parameters of the internal noise source have been measured in the pre-launch thermal tests. They can be further calibrated in the process of in-flight verification by observing a natural celestial body.
The in-flight $T_{\rm sys}$ calibration procedure included the following steps:\\
a) Pointing the main satellite antenna to ``cold sky'' regions; \\
b) Measuring baseline system power levels with the noise source deactivated (OFF); \\
c) Activating the calibration noise source (ON) via dedicated cryogenic receiver coupling port; \\
d) Computing the system noise temperature using differential power measurements.

\begin{figure}[H]
\centering
  \includegraphics[width=\columnwidth]{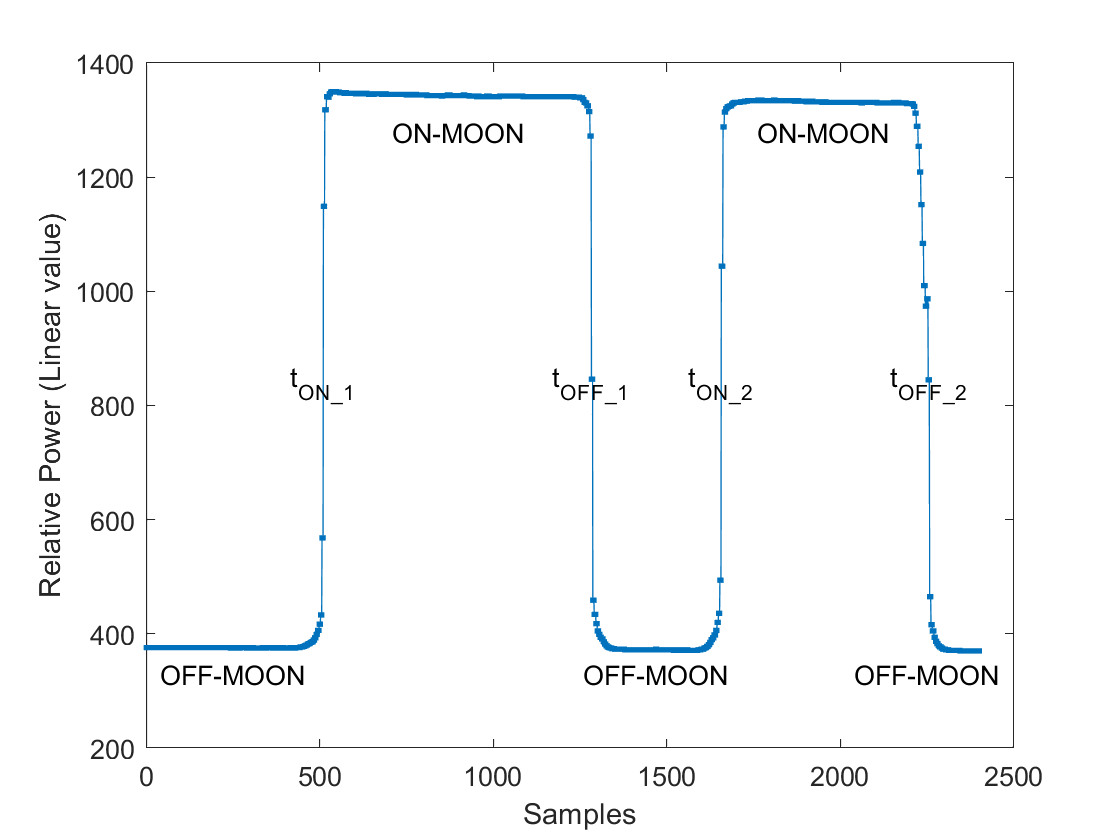}
  \captionsetup{justification=raggedright}
  \caption{Time sequence of measurements of the system temperature of the QueQiao-2 LOVEX instrumentation by intermittent ``single dish'' observations of the Moon and a ``cold sky'' patch. The observation was carried out on 18 July 2024.}
\label{fig:LOVEX-Tsys}
\end{figure}

\begin{figure}[H]
\centering
  \includegraphics[width=\columnwidth]{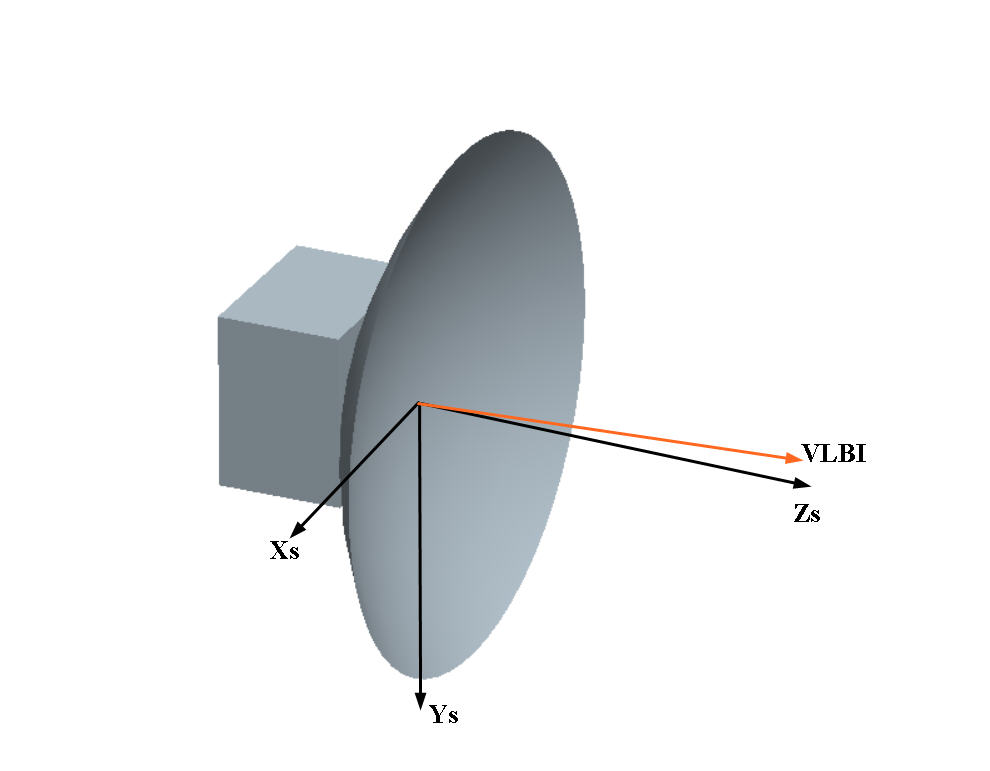}
  \captionsetup{justification=raggedright}
  \caption{A scheme of the 4.2-m antenna rigid mounting on the QueQiao-2 satellite body. The red arrow shows the antenna axis as defined by the position of the VLBI feed. Three black axis show the the body-axis coordinate system of the QueQiao-2 satellite.} 
\label{fig:frame-scheme}
\end{figure}

This approach provides estimates of the system noise temperature by comparing the known injected power of the calibration noise source with the output power of the system in both the `hot' (noise source ON) and `cold' (noise source OFF) conditions. 

The lunar surface, with its well-characterised radio emission properties, served as an additional external calibration source. The angular diameter of the Moon as seen from the QueQiao-2 apolune 
is about 12$^{\circ}$. The primary beam of the 4.2-m antenna at X-band is about 0.6$^{\circ}$. Thus, in observations of the Moon, it covers completely the primary beam as well as the main sidelobes of the antenna. Such the correspondence between the source of known temperature (the Moon) and the primary beam allows us to estimate the system noise temperature by comparing the response to observing a ``cold sky''. The measurement sequence first includes pointing towards a patch of cold sky and measuring the signal power $P_{\rm off}$. Then the antenna points towards the Moon, resulting in detection the signal power $T_{\rm moon}$. The system noise temperature is calculated as 
\begin{equation}
T_{\rm sys}=T_{\rm moon} \cdot P_{\rm off}/(P_{\rm on}-P_{\rm off})  \,\,\, ,
\end{equation} 
\noindent As an input into this estimate of $T_{\rm sys}$ we consider the brightness temperature of the Moon to be $T_{\rm moon}= 235$~K \cite{{Moon-temperature}} . Fig.~\ref{fig:LOVEX-Tsys} shows the sequence of this measurement. The entire process was implemented with two periods towards the Moon, providing four estimates of $T_{sys}$.
The resulting average value of the system temperature is determined to be $\sim 89.3\pm5.0$~K.

\subsubsection{Antenna primary beam and pointing calibration}
\label{sss:ant-beam}

Precise antenna pointing is a critical operational ingredient for successful VLBI operations. Since the 4.2~m antenna is rigidly mounted on the QueQiao-2 satellite, the antenna pointing is adjusted by changing the satellite attitude. Fig.~\ref{fig:frame-scheme} presents a schematic view of the antenna's main axis defined by the position of the VLBI feed (red arrow) in respect to the body-axis coordinate system of the QueQiao-2 satellite (black axis $X_{\rm s}$, $Y_{\rm s}$ and $Z_{\rm s}$). 
The VLBI feed is mounted at a 1.6$^{\circ}$ offset from the communication (relay) feed as mentioned in Section~\ref{ss:queq-2}. After a series of Earth and Moon observations, the offset of the VLBI feed was established in QueQiao-2 frame as $0.143^\circ$ in the YZ plane and $1.604^\circ$ in the XZ plane.

\begin{figure}[H]
\centering
  \includegraphics[width=\columnwidth]{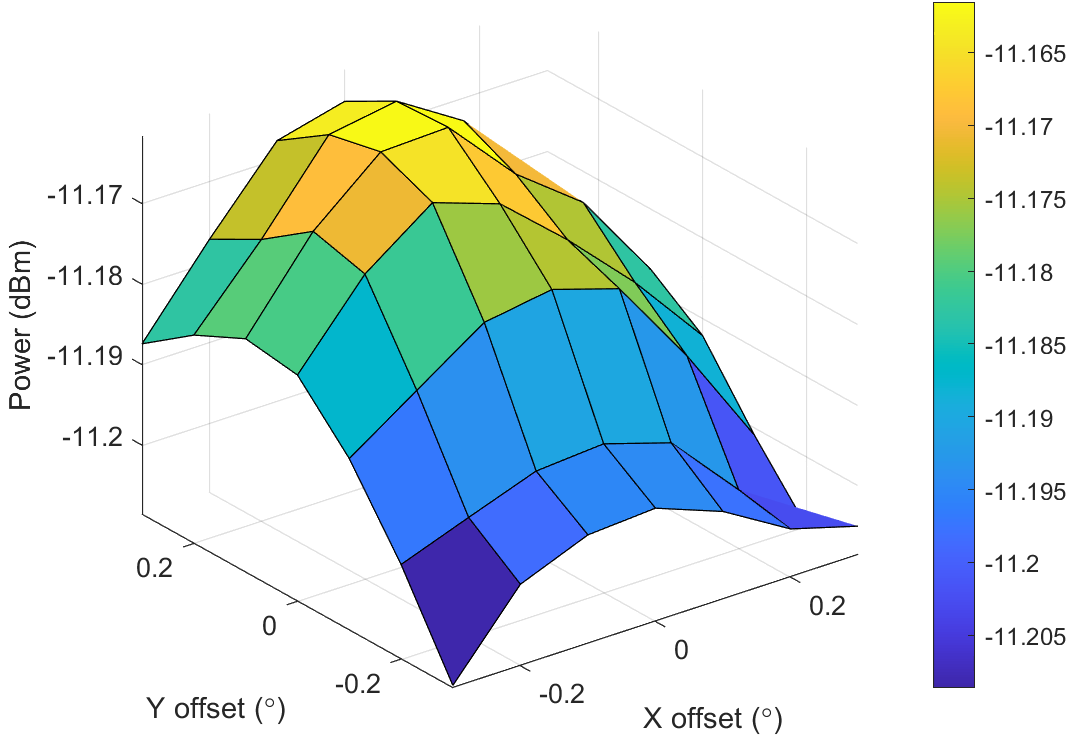}
  \captionsetup{justification=raggedright}
  \caption{A boresighting pattern across a $\pm 0.3$~degree range around the source Tau~A with 0.1~degree step increments along X- and Y-axes for calibration of the antenna primary beam power distribution in the first raster test on 10 September 2024. The vertical axis and colour wedge are graduated in relative power units.} 
\label{fig:rstr}
\end{figure}

In subsequent testing, QueQiao-2 conducted a raster scanning targeting Taurus~A (3C\,144, the Crab Nebula M1). This procedure allowed us to calibrate the QueQiqo-2 VLBI primary beam and its pointing. The choice of this calibration source is based on its stable high flux density, well known continuum spectral properties, and appropriate angular size relative to the QueQiao-2's antenna primary beam. The raster scan sequence was implemented with a spiral pattern across a $\pm 0.3^\circ$ sky patch with $0.1^\circ$ incremental steps along both X and Y axes to map the power distribution pattern, as illustrated in Fig.~\ref{fig:rstr}.

\begin{figure}[H]
\centering
\includegraphics[width=0.95\columnwidth]{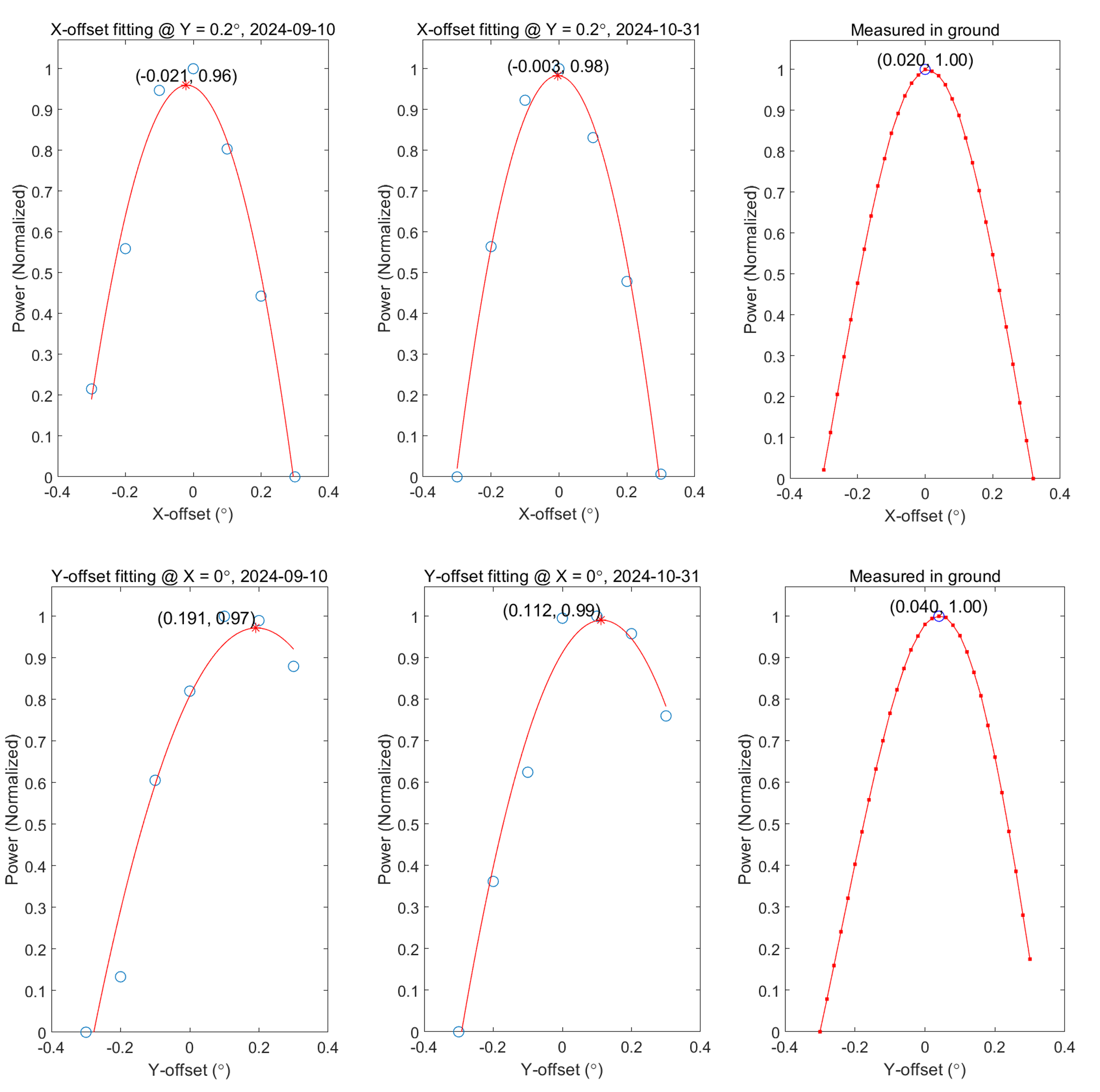}
\captionsetup{justification=raggedright}
\caption{Pointing calibration results. The left and middle figures (top along X axis, and lower along Y axis) were the results obtained on 10 September and 31 October 2024, respectively. The right column plots show the pre-flight laboratory measurements.}
\label{fig:pointing}
\end{figure}

Two pointing calibration campaigns were conducted.  
The first raster test was carried out on 10 September 2024 with the compensation offset of $0.143^\circ$ in the YZ plane and $1.604^\circ$ in the XZ plane. The obtained power distribution is presented in Fig.~\ref{fig:rstr}. Fig.~\ref{fig:pointing} shows the power distributions along the X axis (top left) and Y axis (lower left). The X-axis residual error is less than 0.02~degrees and Y-axis residual error is about 0.19~degrees.

After applying an additional Y-axis correction of $0.06^\circ$, the second raster test was carried out on 31 October 2024 with the compensation of 0.143~degrees in the YZ plane and 1.664~degrees in the XZ plane. Fig.~\ref{fig:pointing} shows the power distributions along X~axis (top middle) and Y~axis, respectively. The X-axis residual error was found to be less than 0.01~degrees and Y-axis residual error about 0.11~degrees. 
\begin{figure}[H]
\centering
  \includegraphics[width=\columnwidth]{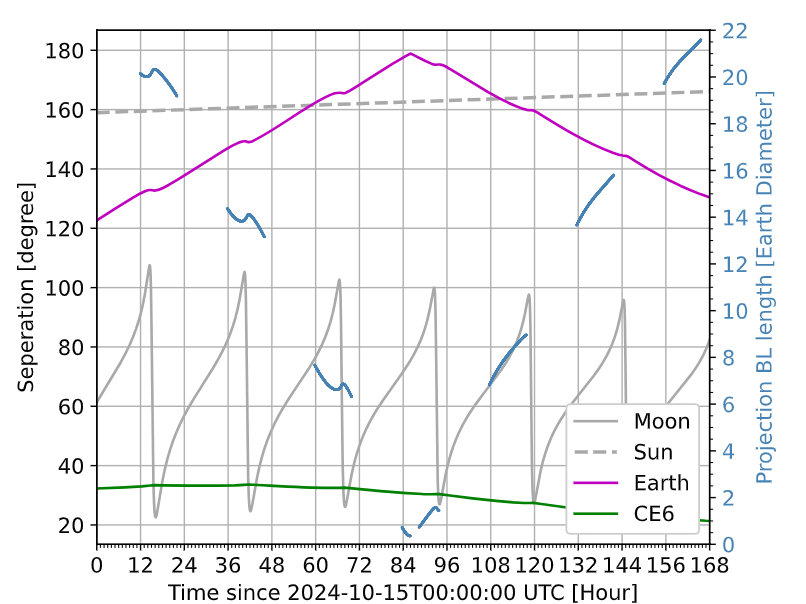}
  \captionsetup{justification=raggedright}
  \caption{Visibility conditions for the first LOVEX VLBI test during the week starting on 15 October 2024.}
\label{fig:tst-vis}
\end{figure}

Fig.~\ref{fig:pointing} also shows the comparative antenna beam patterns in elevation (top right) and azimuth (lower right) directions obtained during the pre-launch laboratory tests, providing valuable performance baselines.  

A series of boresighting tests with the QueQiao-2 antenna, alternating between Earth and Taurus-A, were conducted between July and October 2024. These tests determined the LOVEX feed’s offset to be 0.153 and 1.774 degrees along the $X$ and $Y$ axes of the spacecraft frame (Fig.~\ref{fig:pointing}), respectively. The pointing accuracy (~1$\sigma$) was estimated to be less than 0.1~degrees. Similar antenna pointing calibration observations were occasionally carried out in the following months, for example, on 15 January 2025. We will continue the antenna pointing measurement occasionally for the further calibration. 

Single-dish LOVEX in-flight tests were conducted with 1, 2, 4 and 8 IF channels corresponding to the total bandwidths of 64, 128, 256 and 512~MHz, respectively. 

%
\begin{figure*}
\centering
  \includegraphics[width=0.95\textwidth]{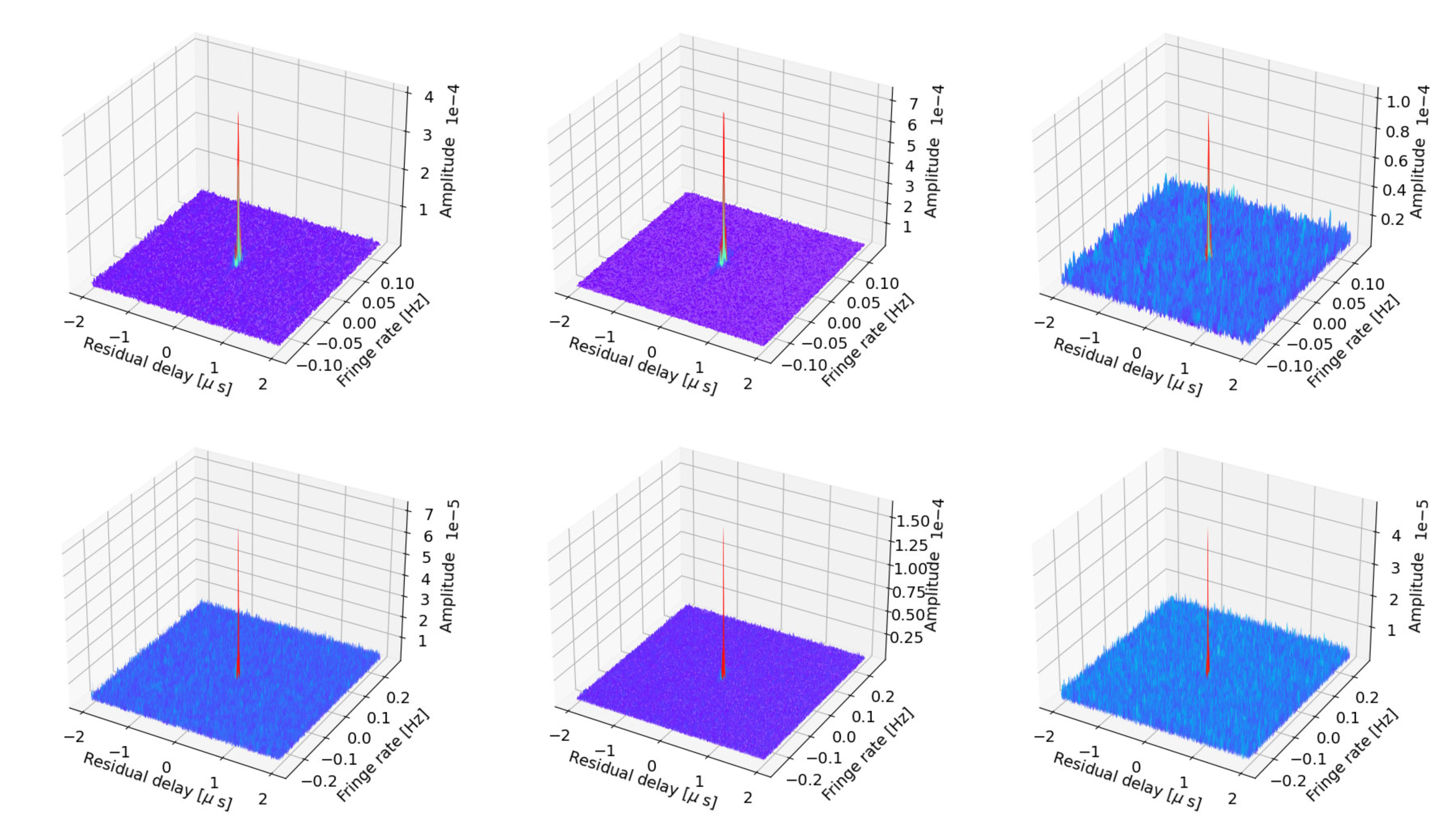}
  \captionsetup{justification=raggedright}
  \caption{Fringe fitting results of the LOVEX observations of the source AO\,0235+164, a total integration time 300~s. The vertical axis shows the normalized amplitude of the output by the correlator. \textit{Top row:} Fringe detections in the observation of 18 October 2024 on the baselines between QueQiao-2 and (left to right) Kunming, Tianma and Nanshan, the bandwidth 64~MHz, the projected baselines $\sim 0.37$~Earth diameters. \textit{Bottom row:} Fringe detections in the observation of 23 January 2025 on the baselines between QueQiao-2 and the same three Earth-based telescopes, the bandwidth 512~MHz, the projected baselines $\sim 5.5$~Earth diameters. }
\label{fig:orb-fri-tsts}
\end{figure*}

\section{Live in-flight LOVEX interferometric tests and operations}
\label{s:svlbi-tst}

Following successful completion of the autonomous in-flight tests by October 2024, both the spaceborne and Earth-based segments of LOVEX have been declared ready for end-to-end SVLBI tests. Table~\ref{tab:coord-st} presents the expected baseline sensitivity values and SNRs for a hypothetical source of 50~mJy flux density with fringe search integration time of 300~s. As known from the practical VLBI experience, such estimates are usually excessively optimistic. So, in order to achieve a guaranteed detection, a choice of the test observing target and observational setup is critical. 
One of the brightest known blazars, AO~0235$+$164, was selected as the observing target in this test (see Section \ref{sss:non-img} for justification). 

The SVLBI testing strategy implies that test observations of AO\,0235$+$164 starts from geometrical configurations that provide shortest projected baselines to maximize fringe detection probability, progressing systematically toward longer projected baselines as system performance has been verified. 

In order to increase the odds of detecting interferometric fringes, the early tests were conducted under the following requirements (in addition to usual for VLBI requirements of common visibility of the target source): \\
-- The Moon does not occult the target for Earth-based telescopes; \\
-- The Sun is at a sufficiently large angular distance from the target; \\ 
-- The Earth is at a sufficiently large angular distance from the target as viewed from QueQiao-2; \\
-- The Moon is at a sufficiently large angular distance from the target as viewed from  QueQiao-2; \\
-- The projected baseline between Earth-based telescopes and QueQiao-2 does not exceed the Earth diameter by more than a factor of $\sim$3. \\
Additional constraints were imposed by operational limitations of the service systems of QueQiao-2 satellite and the Earth-based infrastructure supporting the mission (e.g., data acquisition stations).
Fig.~\ref{fig:tst-vis} illustrates potential parameters of the LOVEX SVLBI test observation of AO~0235$+$164 during one week starting on 15 October 2024. 

\begin{table*}[t]
\footnotesize
\caption{. \,\,\, Comparison of main parameters of VSOP/HALCA, RadioAstron and LOVEX SVLBI systems.
}
\label{tab:svlbi-compa}
\tabcolsep 18pt 
\begin{tabular*}{\textwidth}{lccc}
\toprule
Parameter                    & VSOP/HALCA  & RadioAstron & LOVEX     \\\hline
In-flight operations period  & 1997--2003  & 2011--2019  & 2024--    \\
Antenna diameter [m]         & 8.0         & 10          & 4.2       \\
Observing bands [GHz]        & 1.6, 4.9, 22$^a$ & 0.327, 1.6, 5.0, 22 & 8.4 \\
Longest baseline [$D_{\oplus}$] & 3 & 28 & 33 \\
Polarization (circular)      & single & dual   & dual \\
Maximum registration bandwidth per polarization [MHz] & 32  & 32$^b$ & up to 512 (dual-pol., tunable) \\
Aggregate data rate [Mbps] & 128 & 128 & 2048 \\
\bottomrule
\end{tabular*}
\begin{tablenotes}
   \item[] $^{(a)}$The 22~GHz receiver did not operate in orbit.
   \item[] $^{(b)}$The bandwidth was 16~MHz at the observing band 0.327~GHz.
\end{tablenotes}
\end{table*}
The first VLBI observing test was conducted on 18 October 2024 during the time range 13:30:00 to 13:50:00~UTC. The data were recorded with a bandwidth of 64~MHz (8428-8492~MHz). The ground VLBI stations Tianma, Kunming and Nanshan participated in the experiment. Interferometric fringes were detected on all baselines with a total integration time of 300 seconds. Fig.~\ref{fig:orb-fri-tsts}~(top row) shows the LOVEX interferometric fringes on the baseline between QueQiao-2 (LX) and all three Erth-based telescopes with the projected baseline length of $\sim$0.37 Earth diameters.

The next observation of AO~0235+164 was carried out with the same three Earth-based telescopes on 23 January 2025, 10:20:00 to 10:40:00~UTC, with a bandwidth of 512~MHz (8108--8620~MHz) and baselines to QueQiao-2 $\sim$5.5 Earth diameters. Fringes were detected on all baselines (Fig.~\ref{fig:orb-fri-tsts}, bottom row).

\begin{figure}[H]
\centering
  \includegraphics[width=\columnwidth]{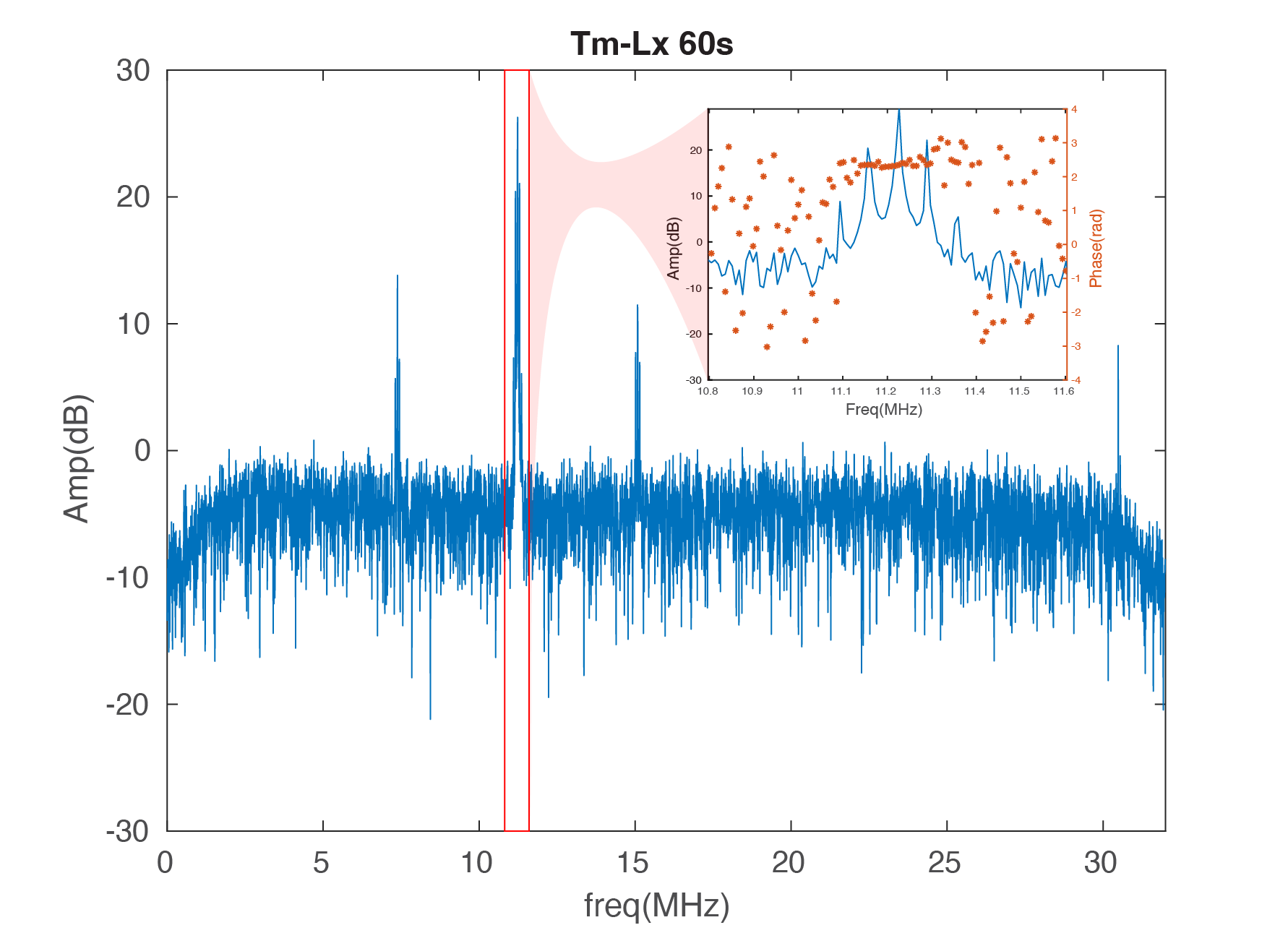} 
  \captionsetup{justification=raggedright}
  \caption{Cross-spectrum obtained in the LOVEX observation of the Chang'E-6 spacecraft on 18 October 2024. The blue line  represent the amplitude of the interferomerc response on the  QueQiao-2 (LX)--Tianma (TM) baseline. 
  The red dots in the top right panel represent the phase of the LX-TM baseline. The projected baseline length is about 12 Earth diameters. The telemetry signal of CE-6 is in $\Delta$DOR format. }
\label{fig:fri-lxtm4Oct18-CE6}
\end{figure}

LOVEX was also tested in the nfVLBI regime. In this test, the target was the Chang'E-6 orbiter. After completing its lunar mission, the orbiter proceeded to the Sun–Earth Lagrangian point L2. On 18 October 2024, LOVEX conducted an nfVLBI observation of the Chang'E-6 orbiter using its X-band $\Delta$DOR signal. The test involved QueQiao-2 and ground telescopes Tianma, Kunming, and Nanshan. Fringes were obtained on all baselines to QueQiao-2. Fig.~\ref{fig:fri-lxtm4Oct18-CE6} shows the cross-spectrum of the main carrier and two sub-carriers, as well as the  amplitude and phase of the interferomerc response on the QueQiao-2 (LX)–Tianma (TM) baseline. Notably, this was the first nfVLBI observation conducted on an Earth–Space baseline.

Based on the SVLBI tests described above, the LOVEX system has been declared fully operational. At the time of this writing, the accumulation of LOVEX observing data was ongoing in preparation for upcoming scientific evaluation.

\section{Conclusions}
\label{s:cncl}

LOVEX is the world’s fourth operational space VLBI mission. Table~\ref{tab:svlbi-compa} presents the main characteristics of LOVEX compared to the two previous dedicated SVLBI missions, VSOP/HALCA and RadioAstron. LOVEX has achieved the following milestones: \\
-- the longest interferometric baseline, reaching approximately 380,000 km; \\
-- the widest observing bandwidth (512 MHz) and highest data rate (2 Gbps) among all Space VLBI systems;\\
-- LOVEX, operating at X-band (8.1--8.6 GHz, the deep space communication frequency band), enabled the first demonstration of near-field VLBI on Space-Earth baselines;  \\
-- the first implementation of onboard storage for raw VLBI data in a space mission context.

In addition, LOVEX has successfully demonstrated the operational viability of a space-qualified passive hydrogen maser frequency standard for high-precision space VLBI timing applications. Furthermore, LOVEX represents a significant achievement as the first fully national Space VLBI system, while all previous SVLBI missions have been based on international efforts.

LOVEX paves the way for the continued development and deployment of advanced Space VLBI systems, which will support a wide range of scientific applications in astrophysics, astrometry, fundamental physics, and planetary science. The technical innovations demonstrated by LOVEX, particularly its integration with the QueQiao-2 lunar relay satellite, showcase an efficient approach to implementing scientific instrumentation by leveraging existing space mission infrastructure.

The successful detection of fringes from both astronomical sources and spacecraft transmissions demonstrates the system's versatility and opens new possibilities for ultra-high-resolution radio astronomy and precise spacecraft tracking. As data collection continues and scientific analysis progresses, LOVEX is expected to provide insights into astrophysical environments and contribute to future deep space exploration capabilities.

\Acknowledgements{
This work was supported by Lunar Orbital VLBI Experiment of the Chang'E-7 Mission in Chinese Lunar Exploration Program. 
Shanghai Astronomical Observatory acknowledges cooperation in developing VLBI-specific payload with Shandong Aerospace Electronics Technology Institute, Xi’an Microelectronic Technology Institute, Shanghai Aerospace Electronics Co. Ltd., Shanghai Aerospace Electronics Technology Institute, and Technical Institute of Physics and Chemistry of CAS.
LIG gratefully acknowledges support by the International Talent Program of the Chinese Academy of Sciences, project number 2024PVA0008.}

\InterestConflict{The authors declare that they have no conflict of interest.}


\end{multicols}

\begin{thebibliography}{99}
\bibitem {TMS-2017} A.~R.~Thompson,  J.~M.~Moran, G.~W.~Swenson,
\textit{Interferometry and Synthesis in Radio Astronomy},  Jr. 3rd ed. (Springer, 2017), doi:10.1007/978-3-319-44431-4

\bibitem{LIG-2020} L.~I.~Gurvits, \textit{Space VLBI: from first ideas to operational missions}, Advances in Space Research 65, 868–-876, (2020), doi:10.1016/j.asr.2019.05.042

\bibitem{LIG-2023} L.~I.~Gurvits, \textit{A Brief History of Space VLBI}, in 8th IEEE History of Electrotechnology Conference (HISTELCON), 171--174, (2023), arXiv:2306.17647. doi:10.48550/arXiv.2306.17647

\bibitem{TDRSS-1986} G.~S.~Levy, R.~P.~Linfield, J.~S.~Ulvestad,  et al.,
\textit{Very Long Baseline
Interferometric Observations Made with an Orbiting Radio Telescope},
Science 234(4773), 187–189, (1986).

\bibitem{VSOP-1998Sci} H.~Hirabayshi, H.~Hirosawa, H.~Kobayashi, et al.,
\textit{Overview and Initial Results of the Very Long Baseline Interferometry Space Observatory Programme},
Science 281(4773), 1825–1829, (1998).

\bibitem{SVLBI-2000} \textit{Astrophysical Phenomena Revealed by Space VLBI}, Proceedings of the VSOP Symposium, eds.: H.~Hirabayashi, P.~G.~Edwards, and D.~W.~Murphy, (Institute of Space and Astronautical Science, 2000).

\bibitem{ASPC-402-2009} \textit{Approaching Micro-Arcsecond Resolution with VSOP-2: Astrophysics and Technologies}, eds.: Y.~Hagiwara, E.~Fomalont, M.~Tsuboi, and Y.~Murata, ASP Conference series 402, (2009).

\bibitem{NSK+2013RA} N.~S.~Kardashev, V.~V.~Khartov, V.~V.~Abramov, et al., \textit{“RadioAstron”–
A Telescope with a Size of 300 000 km: Main Parameters and First Observational Results}, Astronomy Reports 57(3), 153--194, (2013).

\bibitem{Kovalev+2020AdSpR} Y.~Y.~Kovalev, N.~S.~Kardashev,  K.~V.~Sokolovsky, et al., \textit{Detection statistics of the RadioAstron AGN survey}, Advances in Space Research 65, 705--711, (2020), doi:10.1016/j.asr.2019.08.035.

\bibitem{Giovannini+2018NatAs} G.~Giovannini, T.~Savolainen, M.~Orienti, et al., \textit{A wide and collimated radio jet in 3C84 on the scale of a few hundred gravitational radii},  Nature Astronomy 2, 472--477, (2018),  doi:10.1038/s41550-018-0431-2
\bibitem{Fuentes+2023NatAs} A.~Fuentes, J.~L.~G{\'o}mez, J.~M.~Mart{\'\i}, et al., \textit{Filamentary structures as the origin of blazar jet radio variability}, Nature Astronomy 7, 1359, (2023),  doi:10.1038/s41550-023-02105-7

\bibitem{Sobolev+2018IAUS} A.~M.~Sobolev, N.~N.~Shakhvorostova, A.~V.~Alakoz, W.~A.~Baan, \textit{RadioAstron space-VLBI project: studies of masers in star forming regions of our Galaxy and megamasers in external galaxies}, Astrophysical Masers: Unlocking the Mysteries of the Universe, IAUS 336, 417–421, (2018),  doi:10.1017/S1743921317011401.

\bibitem{Smirnova+2014ApJ} T.~V.~Smirnova, V.~I.~Shishov, M.~V.~Popov, et al., \textit{RadioAstron Studies of the Nearby, Turbulent Interstellar Plasma with the Longest Space-Ground Interferometer Baseline}, Astrophysical Journal 786, 115, (2014), doi:10.1088/0004-637X/786/2/115.

\bibitem{Popov+2020ApJ} M.~V.~Popov, N.~Bartel, M.~S.~Burgin, M.~S., et al., \textit{Substructure of Visibility Functions from Scattered Radio Emission of Pulsars through Space VLBI}, Astrophysical Journal 888, 57, (2020), doi:10.3847/1538-4357/ab5db6.

\bibitem{Litvinov+2018PhLA} D.~A.~Litvinov, V.~N.~Rudenko, A.~V.~Alakoz,  et al.,  \textit{Probing the gravitational redshift with an Earth-orbiting satellite}, Physics Letters A 382, 2192–2198, (2018), doi:10.1016/j.physleta.2017.09.014.

\bibitem{Nunes+2023CQGra} N.~V.~Nunes, N.~V., N.~Bartel, A.~Belonenko, et al., 2023, \textit{Gravitational redshift test of EEP with RadioAstron from near Earth to the distance of the Moon}, Classical and Quantum Gravity 40, 175005, (2023), doi:10.1088/1361-6382/ace609.

\bibitem{Hong-2020} X.~Y.~Hong, X.~Z.~Zhang, W.~M.~Zheng, et al.,
\textit{Research progress of VLBI technology and application to China Lunar Exploration Project}, Journal of Deep Space Exploration, 7(4):321-331 (2020)

\bibitem{Dong-2018} G.L.~Dong, H.T.~Li, W.H.~Hao, et al.,
\textit{Development and future of China’s deep space TT\&C system}, Journal of Deep Space Exploration, 5(2), 99-114, (2018)

\bibitem{Qian-Li-2012} Z.~H.~Qian, J.~L.~Li,
\textit{Application of Very Long Baseline Interferometry Technology in Deep Space Exploration}, Science and technology of China press 4, 77, (2012), ISBN 978-7-5046-6048-0

\bibitem{Hong-2014} X.~Y.~Hong, Z.~Q.~Shen, T.~An,
\textit{The Chinese space Millimeter-wavelength VLBI array-A step toward imaging the most compact astronomical objects}, Acta Astronautica, 102, 217, (2014)
\bibitem{An+2020AdSpR} T.~An, X.~Y.~Hong, W.~M.~Zheng, et al., \textit{Space Very Long Baseline Interferometry in China}, Advances in Space Research 65, 850--855, (2020), doi:10.1016/j.asr.2019.03.030.

\bibitem{QQ2} L.~H.~Zhang, L.~Xiong, J.~Sun, et al. 
\textit{System design and validation of QueQiao-2 lunar relay communication satellite},Chinese Space Scienc eand Technology,2024,44(5):23-39

\bibitem{QueQiao-2018} W.~Wu, Y.~Tang, L.~Zhang, D.~Qiao, \textit{Design of communication relay mission for supporting lunar farside soft landing},  Science China Information Sciences, 61(4), 040305, (2018), 
doi:10.1007/s11432-017-9202-1


\bibitem{Zhong+LOVEX-2025} W.~Y.~Zhong, R.~J.~Zhu, Y.~H.~Xie, et al., \textit{Lunar Orbital VLBI Experiment: VLBI-specific Payload Development}, Sci China, Technological Sciences (in review).

\bibitem{ZhangJ+LOVEX-2025} J.~Zhang, W.~Zheng, L.~Liu, et al., \textit{Lunar Orbital VLBI Experiment: Data Correlation System}, Sci China,,Technological Sciences (in review).

\bibitem{Zou+2020} Y.~Zou, Y.~Liu, Y.~Jia, \textit{Overview of China's Upcoming Chang'E Series and the Scientific Objectives and Payloads for Chang'E 7 Mission}, 51st Lunar and Planetary Science Conference, LPI, 1755 (2020)

\bibitem{PRIDE-2023} L.~I.~Gurvits, G.~Cim{\`o}, D.~Dirkx, et al., \textit{Planetary Radio Interferometry and Doppler Experiment (PRIDE) of the JUICE Mission}, Space Sci. Reviews 219, 79, (2023), doi:10.1007/s11214-023-01026-1

\bibitem{Kel-Pau-1969} K.~I.~Kellermann, I.~I.~K.~Pauliny-Toth, \textit{The inverse Compton catastrophe and high brightness
temperature radio sources}, Astropys. J. 155, L71 (1969)

\bibitem{Frey+2000} S.~Frey, L.~.I.~Gurvits, D.~R.~Altschuler, et al., \textit{Dual-Frequency VSOP Observations of AO 0235+164},  Publications of the Astronomical Society of Japan 52, 975–982, (2000), doi:10.1093/pasj/52.6.975

\bibitem{Kovalev+2016-3C273} Y.~Y.~Kovalev, N.~S.~Kardashev, K.~I.~kellermann, et al., \textit{RadioAstron Observations of the Quasar 3C273: A Challenge to the Brightness Temperature Limit}, Astrophys. J. 820, L9, (2016), doi:10.3847/2041-8205/820/1/L9

\bibitem{Johnson+2016-3C273} M.~D.~Johnson, Y.~Y.~Kovalev, C.~R.~Gwinn, et al., \textit{Extreme Brightness Temperatures and Refractive Substructure in 3C273 with RadioAstron}, Astrophys. J. 820, L10, (2016), doi:10.3847/2041-8205/820/1/L10

\bibitem{Rees-1967-TB} M.~J.~Rees, \textit{Studies in radio source structure -- I. A relativistically expanding model for variable quasi-stellar radio sources}, MNRAS 135, 345-360, (1967),  doi:10.1093/mnras/135.4.345
\bibitem{Kardashev-2000} N.~S.~Kardashev, \textit{Radio Synchrotron Emission by Protons and Electrons in Pulsars and the Nuclei of Quasars}, Astronomy Reports 44, 719–724, (2000), doi:10.1134/1.1320497

\bibitem{Begelman+2005} M.~C.~Begelman, R.~E.~Ergun, M.~J.~Rees, \textit{Cyclotron Maser Emission from Blazar Jets?} Astrophys. J. 625, 51–59, (2005), doi:10.1086/429550

\bibitem{Tsang+Kirk-2007} O.~Tsang, J.~G.~Kirk, \textit{The inverse Compton catastrophe and high brightness temperature radio sources}, Astronomy and Astrophysics 463, 145–152, (2007),  doi:10.1051/0004-6361:20066502

\bibitem{Tingay+2003var} S.~J.~Tingay, D.~L.~Jauncey, E.~A.~King, et al., \textit{ATCA Monitoring Observations of 202 Compact Radio Sources in Support of the VSOP AGN Survey}, Publications of the Astronomical Society of Japan 55, 351–384, (2003), doi:10.1093/pasj/55.2.351

\bibitem{Liu+2018var} J.~Liu, H.~Bignall, T.~P.~Krichbaum, et al.,  \textit{Effelsberg Monitoring of a Sample of RadioAstron Blazars: Analysis of Intra-Day Variability},  Galaxies , 49, (2018),  doi:10.3390/galaxies6020049

\bibitem{Hirabayashi+2000PASJ} H.~Hirabayashi, E.B.~Fomalont, S.~Horiuchi, et al., \textit{The VSOP 5 GHz AGN Survey I. Compilation and Observations}, Publ. Astronomical Soc. of Japan 52, 997–L1014 (2000), doi:10.1093/pasj/52.6.997


\bibitem{Cho+2024} I.~Cho, J.~L.~G{\'o}mez, R.~Lico, et al., \textit{Unveiling the bent-jet structure and polarization of OJ 287 at 1.7 GHz with space VLBI}, A\&A, 683, A248, (2024), doi:10.1051/0004-6361/202347157

\bibitem{Krasna+2025} H.~Kr{\'a}sn{\'a}, C.S.~Jacobs, M.~Schartner, P.~Charlot, \textit{A celestial reference frame derived from observations with the Very Long Baseline Interferometry Global Observing System}, A\&A, 693, A16, (2025), doi:10.1051/0004-6361/202451996

\bibitem{Will}C.~M.~Will, \textit{The Confrontation between General Relativity and Experiment}, Living Reviews in Relativity, 17, 4 (2014)

\bibitem{Fomalont}E.~Fomalont, S.~Kopeikin, G.~Lanyi, and J.~Benson, \textit{Progress in measurements of the gravitational bending of radio waves using the VLBA}, Astrophys. J., 699, 1395 (2009) 

\bibitem{Lambert} S. B.~Lambert and C.~Le Poncin-Lafitte, \textit{Improved determination of $\gamma$ by VLBI}, Astron. Astro- phys., 529, A70 (2011)

\bibitem{Bertotti} B.~Bertotti, L.~Iess,  and P.~Tortora, \textit{A test of general relativity using radio links with the Cassini spacecraft}, Nature, 425, 374–376 (2003)

\bibitem{Liu+2022} Q.~Liu, Y.~Huang, F.~Shu, et al., \textit{VLBI technique for the orbit determination of Tianwen-1}, Scientia Sinica Physica, Mechanica \& Astronomica, 52, 239507, (2022), doi:10.1360/SSPMA-2021-0204

\bibitem{Zhang+2024TW} H.~Zhang, F.~Li, L.~Meng L., et al., \textit{Characteristics and close-range exploration methods of near-Earth asteroid 2016HO$_{3}$.}, Astronomical Techniques and Instruments, 1, 42, (2024), doi:10.61977/ati2024004

\bibitem{Moon-temperature}D.~D.~Morabito, W.~Imbriale, and S.~Keihm, \textit{Observing the Moon at Microwave Frequencies Using
a Large-Diameter Deep Space Network Antenna}, IEEE TRANSACTIONS ON ANTENNAS AND PROPAGATION, VOL. 56, NO. 3, 650, (2008)


\end{thebibliography}
\end{document}